\documentclass[12pt,twoside,english,a4paper]{article}
\usepackage{caption}
\usepackage[utf8]{inputenc}
\usepackage[intlimits] {amsmath}   
\usepackage{graphicx}
\usepackage{psfrag}
\usepackage{ifthen}
\usepackage{fancyhdr}
\usepackage{rotating}
\usepackage{multirow}
\usepackage{booktabs}              
\usepackage{amssymb}
\usepackage{amsbsy}
\usepackage{bm}                    
\usepackage{bbm}
\usepackage{babel}
\usepackage{theorem}
\usepackage{upgreek}  
\usepackage{pifont}   
\usepackage[active]{srcltx}   

\usepackage{tikz}
\usepackage{pgfplots}
\usepackage{nicefrac}
\usepackage[ruled]{algorithm2e}
\usepackage{appendix}
\usepackage{longtable}
\usepackage{url}
\usepackage{graphicx}
\usepackage[
    colorlinks=false,
    citebordercolor={cyan}
]{hyperref}
\usepackage{xcolor}
\usepackage{bm}
\usepackage{tabularray}
\usepackage{adjustbox}
\usepackage{transparent}
\usepackage{subcaption}
\usepackage{float}

\renewcommand{\vec}[1]{\bm{#1}}
\newcommand{\ten}[1]{\bm{#1}}

\usetikzlibrary{positioning, calc, shapes.misc, external}
\pgfplotsset{compat=1.18}

\usepackage{suetterl}  
\newfont{\suetdbl}{suet14 scaled 2000}  

\usepackage{yfonts}  
\newfont{\gothdbl}{ygoth scaled 2000} 
\newfont{\frakdbl}{yfrak scaled 2000}
\newfont{\swabdbl}{yswab scaled 2000}

\usepackage[normalem]{ulem}  
\usepackage[textsize=footnotesize,textwidth=1.7cm]{todonotes}
\usepackage[]{defcommandsls}      

\usepackage[numbers,sort]{natbib}
\fancypagestyle{plain}{%
  \fancyhead{}%
  \fancyfoot[c]{\sffamily\thepage}%
}
\makeatletter                           
\def\cleardoublepage{\clearpage\if@twoside \ifodd\c@page\else
  \hbox{}
  \vspace*{\fill}
  \thispagestyle{empty}
  \newpage
  \if@twocolumn\hbox{}\newpage\fi\fi\fi}
\makeatother
\begin{document}
\unitlength1.0cm
\frenchspacing



\thispagestyle{empty}
{\bf \large
\begin{center}
    Efficient strain-space hyperreduction in large-deformation solid mechanics
\end{center}
}


\vspace{4mm}
\begin{center}
{Erik Faust \& Lisa Scheunemann}
\end{center}

\vspace{4mm}
\begin{center}{Institute of Applied Mechanics (IFAM)}
{RWTH Aachen University, Mies-van-der-Rohe-Str. 1, 52074 Aachen,
Germany}
{\\Chair of Applied Mechanics (LTM)}
{RPTU Kaiserslautern-Landau, Gottlieb-Daimler-Str.,
67663 Kaiserslautern,
Germany}
{\\ \small e-mails: erik.faust@ifam.rwth-aachen.de, lisa.scheunemann@ifam.rwth-aachen.de
    }
    \end{center}

\vspace{4mm}
\begin{center}
{\bf \large Abstract}
\bigskip

{\footnotesize
\begin{minipage}{14.5cm}
\noindent

Strain-space model order reduction (MOR) techniques have recently been shown to achieve exceptional performance in terms of the tradeoff between runtime and accuracy achieved in computational homogenisation problems.
In this article, we generalise such techniques to problems in large-deformation solid mechanics beyond the context of computational homogenisation. Arbitrary-valued, parameterised Dirichlet boundary conditions are satisfied by construction using a lifting with boundary-consistent fields computed offline.
This allows us to pose a version of the Empirical Cubature Method (ECM)~\cite{HerCaiFer:2017:dhn,Her:2020:MMP} in strain space and generalise the Empirically Corrected Cluster Cubature (E3C)~\cite{WulHau:2025:ech,wul:2025:ECC,WulHau:2025:EHR} as well as Empirical Material Sampling and Linearisation (EMSL)~\cite{FauSch:2026:EMS} beyond computational homogenisation problems.

The strain-space versions of EMSL, ECM, and E3C are compared against each other and a standard displacement-space formulation of Energy Conserving Weighting and Sampling (ECSW)~\cite{FarAveCha:2014:drn}. On two hyperelastic example problems with parameterised material behaviour and deformation, the strain-space methods outperform the displacement-space alternative in the tradeoff between runtime and accuracy. E3C and EMSL in particular facilitate 10,000 and 100,000-fold speedups, respectively, while retaining high levels of accuracy. EMSL is shown to be the method of choice when online and offline runtime budgets are very limited, while E3C yields exceptional levels of accuracy when slightly more runtime is acceptable.

\end{minipage}
}
\end{center}


\section{Introduction}

In the context of parameterised solid-mechanical problems, strain-space model order reduction (MOR) techniques are attractive because of their simplicity and low intrusiveness. Element-level operations are bypassed entirely, and only displacement-, strain- and stress data, which can easily be obtained from black-box solvers, are required for training. There is no need to gather snapshots of the residual vector, and details about the discretisation, such as the element types used, need not be available. In addition to the training data and boundary condition information, we only require access to an appropriate material model to deploy a strain-space MOR method to a previously unseen family of problems. If strict non-intrusiveness is desired, a material model could also be inferred from stress-strain pairs.

To the best of our knowledge, however, strain-space MOR techniques have so far only been applied to simulations on representative volume elements (RVEs) in the context of computational homogenisation. In this field, strain-space approaches could roughly be grouped into two loose families: the first includes structure-preserving incremental techniques such as the Nonuniform Transformation field Analysis (NFTA)~\cite{MicSuq:2003:ntf}, Potential-Based Reduced Basis MOR (pRBMOR)~\cite{FriLeu:2013:rbh}, and Self-Consistent Clustering Analysis (SCCA)~\cite{LiuBesLiu:2016:sca,YUKafLiu:2019:SCC}. The second group, which has been explored less extensively to date and will be the focus of this work, contains more general reduced Galerkin approaches such as
Kunc and Fritzen's strain Galerkin scheme~\cite{FriKun:2018:tdh,KunFri:2019:fsh} and the Empirically Corrected Cluster Cubature (E3C) by Wulfinghoff et al.~\cite{WulHau:2025:ech,wul:2025:ECC,WulHau:2025:EHR}. The cluster-enhanced potential-based procedure from~\cite{RuaJuChe:2024:CPR} lies somewhere between these two poles. 

As discussed in our recent pre-print~\cite{FauSch:2026:EMS}, other popular hyperreduced Galerkin techniques such as Energy Conserving Sampling and Weighting (ECSW)~\cite{FarAveCha:2014:drn} or the Empirical Cubature Method (ECM)~\cite{HerCaiFer:2017:dhn,Her:2020:MMP}, can also be reformulated in strain space.
The main point of that work, however, was the introduction of another strain-space MOR technique for RVE problems, called Empirical Material Sampling and Linearisation (EMSL)~\cite{FauSch:2026:EMS}. 
Rather than approximately integrating an exact material law, as E3C does, EMSL exactly integrates a cluster-wise linear approximation of the stress response around empirically estimated reference strains. Because this results in an affine problem in each load step and expensive operations can easily be preprocessed, EMSL yielded significant runtime accelerations in an example large-deformation RVE problem.

Despite their conceptual attractiveness, efficiency, and their success when applied to computational homogenisation, there do not seem to be strain-space MOR approaches for the broader class of three-dimensional solid-mechanical problems. Authors have reported success in exploiting strain-space bases for large-deformation beam problems, as e.g. in~\cite{AbdAnuFel:2025:SSM}. Additionally, it is worth mentioning mixed variational FE formulations based on a Hu–Washizu-type functional, which introduce an additional unknown field for the strain; see e.g.~\cite{FahYav:2019:CMF}. When using such mixed formulations to model shells, reduced models based on truncated Taylor expansions have been used for post-buckling load path following, as e.g. in ~\cite{MagLiaGar:2018:EMV}. Projection-based strain-space MOR techniques for three-dimensional problems with arbitrary boundary values, however, remain to be explored.

At first glance, this appears unsurprising: the performance of E3C and EMSL, as well as strain-space variants of ECM and ECSW, results from operating directly on strains and at integration points. The quantities of interest in most solid-mechanical problems, however, are displacements at nodes. Furthermore, these problems are generally subject to Dirichlet boundary conditions (BCs) specifying displacement boundary values, not strains. Additionally, the integration point strain values appearing in classical numerical schemes like FE are not available at the boundary. In strain space, we can therefore not impose boundary conditions by enforcing boundary values at boundary degrees of freedom, as we can in the case of displacement-space techniques.

In this work, we propose a methodology to circumvent these obstacles. To satisfy nonzero Dirichlet BCs, we use a lifted formulation: BC-consistent strain fields are computed for boundary values of $1$ at each independent boundary. These can be scaled with arbitrary boundary values for automatic BC fulfillment. In analogy to the handling of periodic BCs in RVEs~\cite{Sch:2014:nth}, we then solve for a fluctuation or difference field defined as the difference between the actual strain and the BC-consistent strain field. We use the term "fluctuation field" in the remainder of this work, to stress this similarity with the RVE case. The fluctuation vanishes at all Dirichlet BCs by construction, such that POD modes fulfill all BCs a priori. This boundary handling approach is closely related to the lifting function approach used in some POD-Galerkin models of fluid flows, see e.g.~\cite{StaHijMol:2017:PGR,BusStaRoz:2020:PGR,GirQuaRoy:2021:PGR}.
Additionally, displacement field results can be postprocessed from the reduced variables.

After introducing the parameterised solid-mechanical target problem class in Subsection~\ref{s:displ}, we discuss boundary handling in Subsections~\ref{s:bc} and~\ref{s:bcnew}. We recapitulate the popular displacement-space POD-ECSW approach in Section~\ref{s:PODECSW}, before moving on to strain-space Galerkin reduction in Section~\ref{s:strain_MOR}.
In Subsection~\ref{s:ECM}, ECM is reformulated in strain space, and E3C is generalised to problems with arbitrary Dirichlet BCs in Subsection~\ref{s:E3C}. We similarly extend EMSL in Section~\ref{s:EMSL}. All methods are applied to two hyperelastic large-deformation problems motivated by inverse parameter estimation in Section~\ref{s:application}: a plate with two holes, which is designed to exhibit diverse strain- and stress states~\cite{FlaKumDel:2022:DPM}, is parameterised in terms of its material behaviour and the extent of loading. We pose a simpler version of this problem with only tensile loading and relatively small deformations, as well as a challenging version subjected to tensile, shear, and mixed load paths and larger deformations. The latter also necessitates a finer mesh and leads to higher computational costs.
We evaluate the performance of classical displacement-space ECSW, strain-space ECM and E3C, as well as EMSL, in terms of the tradeoff between validation accuracy and online as well as offline runtime.

\section{Large-deformation hyperelasticity}

Before delving into the MOR methods to be explored in this work, we summarise essentials of hyperelastic solid mechanics at large deformations in Subsection~\ref{s:displ}. More details can be found in the reference literature~\cite{Wri:2008:nfe,Hol:2002:nsm}, on which descriptions here are also based. We then discuss the handling of boundaries in Subsections~\ref{s:bc} and~\ref{s:bcnew} in some detail, because this implementational aspect will become crucial to the strain-space MOR techniques described in the following.

\subsection{Displacement-space formulation}\label{s:displ}

For a quasi-static large-deformation solid-mechanical problem on domain $\Omega$ with points $\vec{X}\in \Omega$, we can write the weak form of the balance of linear momentum as 
\begin{equation}
    \int_{\Omega} \delta \vec{F}(\vec{X}):\vec{P}(\vec{X}) d\text{V} 
    = 0\,,\label{eq:weakform}
\end{equation}
where $\delta \vec{F}(\vec{X})$ denotes a kinematically admissible perturbation of the deformation gradient $\ten{F}$. 
In the case of hyperelasticity, the first Piola-Kirchhoff stress tensor $\ten{P}$ can be derived from a potential $W$ as $\vec{P}=\frac{\partial W}{\partial \vec{F}}$, while the nominal stiffness is given by $\ten{\mathbb{{A}}}=\frac{\partial\vec{P}}{\partial\vec{F}}$. In this work, we use a Mooney-Rivlin material model
\begin{equation}
    W = c_1(\bar{I}_1-3)+c_2(\bar{I}_2-3) + \frac{\kappa}{4}(J^2-1-2\ln J)\,,\label{eq:MR}
\end{equation}
which is given in terms of the volumetric strain $J=\det \ten{F}$ as well as the right Cauchy-Green tensor $\ten{C}=\ten{F}^T\vec{F}$ and its isochoric invariants $\bar{I}_1=J^{-2/3}I_1$, $\bar{I}_2 = J^{-4/3} I_2$. The shear and bulk moduli can be computed as $\mu=2(c_1+c_2)$ and $\kappa = \lambda+\frac{2}{3}\mu$, respectively. Since bulk loads and Neumann boundary conditions are straightforward to enforce in reduced models, we do not discuss these in detail here. The focus of this work are the more challenging Dirichlet BCs.

A standard isoparametric Galerkin Ansatz for the displacement $\vec{v}^{e}(\vec{X})=\sum_k N^{ek}(\vec{X}) \vec{u}^{ek}$ over a set of elements ${e}\in E$ allows us to discretise the weak form in terms of element residuals as
\begin{equation}
    \sum_{{e}\in E} \delta \vec{u}^{{e}kT} \vec{g}^{{e}k}=0,\quad \text{with}\quad \vec{g}^{{e}k} = \int_{\Omega^{e}} \ten{P}\frac{\partial N^{{e}k}}{\partial \vec{X}} \text{d}V  \,,\label{eq:elemres}
\end{equation}
where $N^{ek}(\vec{X})$ denotes the shape function associated with node $k$ in element $e$, and $\delta \vec{u}^{{e}k}$ is the variation of the corresponding displacement. Gathering nodes which are duplicated across elements allows us to assemble a global residual, which can be written either in index or matrix form as
\begin{equation}
    g_{3K+\alpha} = \sum_{\text{e} \in E} g_\alpha^{\text{e}k}\,,\quad \text{or} \quad \vec{g}=\sum_{e \in E} \ten{S}^e \vec{g}^e\,,
\end{equation}
where $\ten{S}^e$ is a sparse sorting matrix. Of course, a practical implementation must be performed in an entry-wise manner, but the matrix formulation will be used below for simplicity of exposition. We can finally write the discretised weak form, which defines the global nonlinear equation system to be solved, as
\begin{equation}
    \delta \Vec{{u}}^T \Vec{g}(\Vec{{u}};\Vec{p}) = 0\,,\quad \text{and thus} \quad  \Vec{g}(\Vec{{u}};\Vec{p}) = \vec{0}\,.\label{eq:eqsystem}
\end{equation}
The parameter vector $\vec{p}$ can e.g. contain loading or material information. In the example explored in this work, we have
\begin{equation}
    \vec{p} = (c_1,c_2,\lambda,d_0,d_1,...)^T\,,
\end{equation}
with prescribed boundary values $d_j$ on some part of the boundary $\partial \Omega_j$. For some parameter value $\vec{p}$, Eq.~\eqref{eq:eqsystem} can be solved iteratively using a Newton-Raphson scheme
\begin{equation}
    \ten{K}(\vec{u_\text{cur}};\Vec{p}) \Delta {\Vec{u}} = - \Vec{g} (\vec{u_\text{cur}};\Vec{p})\,. \label{eq:newton}
\end{equation}
The stiffness matrix $\ten{K}=\frac{\partial \vec{g}}{\partial\vec{u}}$ can also be assembled from element values as 
\begin{equation}
    \ten{K} = \sum_{e \in E} \ten{S}^e \ten{K}^e \ten{S}^{eT}\,, \quad \text{with}\quad {K}_{\alpha\gamma}^{\text{e}kl} = \int_{\Omega^\text{e}} \frac{\partial N^{\text{e}l}}{\partial X_\beta} \mathbb{{A}}_{\alpha\beta\gamma\delta}  \frac{\partial N^{\text{e}k}}{\partial X_\delta} \text{d}V\,,\label{eq:K}
\end{equation}
where index notation and Einstein summation convention are used. 

\subsection{Classical boundary handling techniques}\label{s:bc}

Dirichlet boundary conditions prescribe boundary values $u_j$ on some part of the boundary $\partial \Omega_j$. In the discrete setting, we can denote the boundary degrees of freedom for the $j$-th boundary as $b_j$, and write
\begin{equation}
    \vec{u}_{b_j} \stackrel{!}{=} \vec{1}d_j\,,\quad j\in \{0,...,B\}\,,
\end{equation}
where $\vec{1}\in R^{|b_j|}$ is a vector of ones. In hyperelasticity, known boundary values $d_j$ can be written into $\vec{u}_{b_j}$ directly, and Eq.~\eqref{eq:newton} solved only for the independent degrees of freedom $\vec{u}_i$, i.e.
\begin{equation}
    \ten{K}_{ii}(\vec{u_\text{cur}};\Vec{p}) \Delta {\Vec{u}}_i = - \Vec{g}_i (\vec{u_\text{cur}};\Vec{p})\,,\quad \text{where} \quad \vec{u_\text{cur}}=[\vec{u}_{i,\text{cur}}^T,\vec{u}_{b}^T]^T
\end{equation}
with $\vec{u}_b = [\vec{u}_{b_j}]_{j\in [1,B]}$. Here, we tacitly assume degrees of freedom to have been re-sorted so that boundary degrees of freedom appear at the end of the unknown vector $\vec{u}_\text{cur}$. $\Vec{g}_i$ denotes the degrees of freedom of the residual corresponding to independent nodes only~\cite{Zoh:2018:FEP}. Similarly, $\ten{K}_{ii}$ refers to the block of the stiffness matrix corresponding to independent degrees of freedom. Homogeneous (i.e., zero) Dirichlet BCs, which are a special case, are generally trivial and can just be removed from the equation system. 

\begin{figure}
    \centering
    \def\svgwidth{\linewidth}
    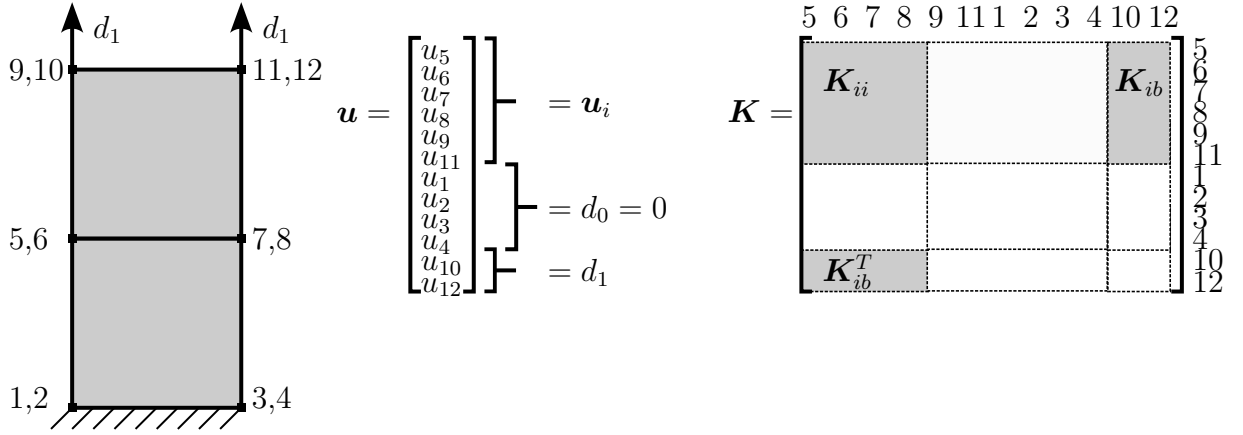
    \caption{Illustration of boundary handling on a minimal example. Degrees of freedom $1-4$ are subject to homogeneous Dirichlet boundary conditions ($j=0$, $d_j = 0$, $b_0 =\{1,2,3,4\}$). Degrees of freedom $10$ and $12$ ($b_1=\{10,12\}$) are subject to nonzero Dirichlet boundary conditions with boundary value $d_1$. The remaining degrees of freedom are independent. Degrees of freedom have been re-sorted to simplify the illustration.}
    \label{fig:BCvis}
\end{figure}

Alternatively and more generally, known displacement increments can be written into $\Delta \vec{u}_b$ and the residual modified, such that
\begin{equation}
     \ten{K}_{ii}(\vec{u_\text{cur}};\Vec{p}) \Delta {\Vec{u}}_i = - \Vec{g}_i (\vec{u_\text{cur}};\Vec{p}) - \ten{K}_{ib}(\vec{u_\text{cur}};\Vec{p}) \Delta {\Vec{u}}_b \,,
\end{equation}
where $\ten{K}_{ib}$ is the block of the stiffness matrix containing rows corresponding to independent and columns corresponding to boundary degrees of freedom.
Rather than incrementing only select degrees of freedom in response to a boundary value change, the effect of the increment is now applied more evenly as though it were an external force. This can improve the conditioning of the equation system and accelerate convergence, especially for history-dependent problems~\cite{BorCriRemVer:2012:nfe}. Fig.~\ref{fig:BCvis} illustrates the relevant blocks of the stiffness matrix and unknown vector by way of a minimal example.

\subsection{Lifting of inhomogeneous Dirichlet boundary conditions}\label{s:bcnew}

Alternatively, we can employ a lifting strategy analogous to that applied to handle periodic BCs in computational homogenisation: we use a reference field which is consistent with BCs a priori, and only compute the fluctuation from this field in our solver~\cite{Sch:2014:nth,Sch:2000:hnk,SaeSteJav:2016:ach}. This transforms all inhomogenous (i.e. nonzero) Dirichlet BCs into homogeneous BCs, and is closely related to the lifting function approach employed by some POD-Galerkin models of fluid flow~\cite{StaHijMol:2017:PGR,BusStaRoz:2020:PGR,GirQuaRoy:2021:PGR}. In the context of full-order displacement-space FE modeling or displacement-space MOR, a lifting like this is not necessary. It is, however, a crucial ingredient to the successful application of strain-space MOR techniques, because we can not simply impose displacement boundary values on strains at integration points which do not lie on the boundary. To illustrate the principle, we first discuss the lifting approach in the full-order FE setting and in displacement space, before moving on to reduced modeling and strain-space approaches in Sections~\ref{s:PODECSW} and~\ref{s:strain_MOR}.

We therefore define a BC-consistent field $\vec{\bar{u}}_j$ for each boundary $j$. This field corresponds to the displacement field solution when degrees of freedom on boundary $j$ are assigned a unit displacement, while all boundary values are set to zero. When scaled with $d_j$, $\vec{\bar{u}}_j$ satisfies BC $j$ by construction (and is zero for all others)\footnote{Where large deformations are expected, it can be helpful to compute $\vec{\bar{u}}_j^*$ for larger boundary values $d_j^*$, and to scale the results, using $\vec{\bar{u}}_j=\frac{1}{d_j^*}\vec{\bar{u}}_j^*$ as a BC-consistent field.}. Given some arbitrary boundary values $d_j$, we can then solve for a displacement fluctuation
\begin{equation}
    \vec{\tilde{u}} = \vec{u} - \sum_{j=1}^B d_j \vec{\bar{u}}_j\,,
\end{equation}
which is zero at all boundaries, such that $\vec{u}$ satisfies boundary conditions a priori. For convenience, we can write this in matrix form as 
\begin{equation}
    \vec{\tilde{u}} = \vec{u} - \ten{\bar{U}} \vec{d}\,,\label{eq:fluc}
\end{equation}
with $\ten{\bar{U}}= [\vec{\bar{u}}_j]_{j\in [1,B]}$, $\vec{d}= [d_j]_{j\in [1,B]}$.
The BC-consistent fields $\vec{\bar{u}}_j$ can for example be obtained via a solution to the parameterised problem in Eq.~\eqref{eq:eqsystem} with
\begin{align*}
    & c_1 = \bar{c}_1\,,\quad c_2 =\bar{c}_2\,, \quad \lambda = \bar{\lambda}\,, \\ 
    & d_i = 0\,, i\neq j\,, \\
    & d_i = 1\,,i= j\,.
\end{align*}
at some reference material parameters $\bar{c}_1,\bar{c}_2,\bar{\lambda}$.
In a full-order model, these reference parameters do not actually influence the solutions identified by the iterative solver in a well-posed problem.
Nevertheless, it makes sense for the reference parameters to lie close to the centre of the parameter space to be explored in the parameterised problem in question, because this improves the initial guesses provided to the nonlinear solver scheme and can accelerate convergence.
In the context of MOR, a clever choice of reference parameters also promotes basis quality (see Section~\ref{s:PODECSW}).
Alternatively, an a priori, purely kinematic lifting could be used. We would like to emphasise that, in our strain-space case, it is not sufficient to simply set the values of the BC-consistent field at the boundary to one and the rest of the field to zero. As explained in Section~\ref{s:strain_MOR}, this would lead to pathological gradients in the BC-consistent strain field. Since the computation of $\vec{\bar{u}}_j$ using reference parameters worked well in the examples we explored, a detailed study about alternative lifting strategies is beyond the scope of this work. 

\begin{figure}
    \centering
    \def\svgwidth{0.95\linewidth}
    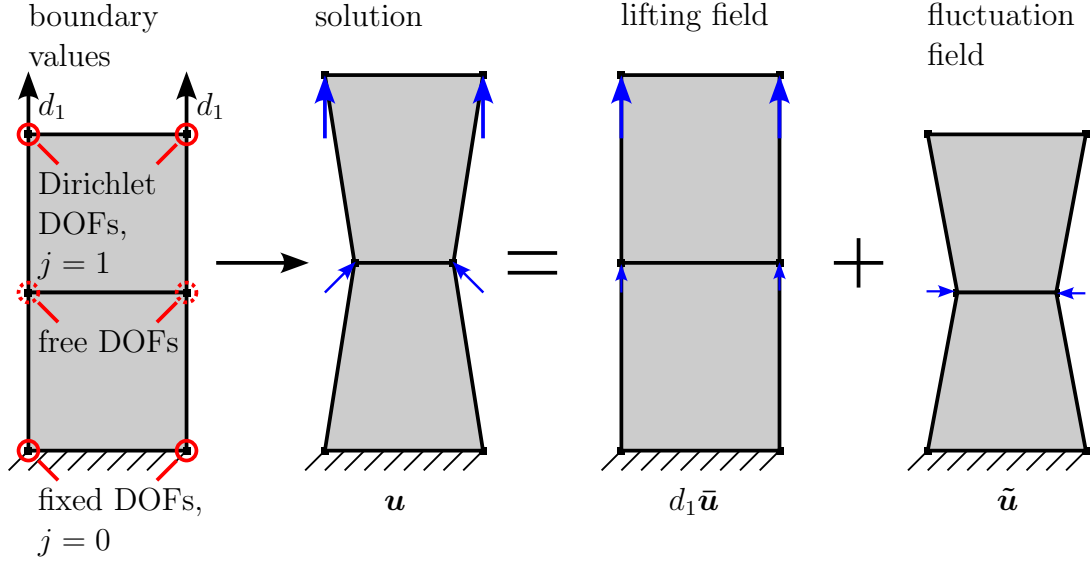
    \caption{Illustration of lifting for inhomogeneous Dirichlet boundary conditions on a minimal example. In this case, the boundary condition field might have been computed for an unlikely material parameter combination with $\nu=0$. Homogeneous and inhomogeneous Dirichlet degrees of freedom, as well as free degrees of freedom are circled and labeled. The solution field is decomposed into the lifting field and the fluctuation field. Displacement vectors at all nodes are shown as blue arrows.}
    \label{fig:BCflucvis}
\end{figure}

Fig.~\ref{fig:BCflucvis} illustrates the boundary handling strategy using the minimal example already introduced in Fig.~\ref{fig:BCvis}. In this case, the BC-consistent field might have been obtained from a simulation with $\nu=0$. For a more realistic example, consider Fig.~\ref{fig:BC}. There, a BC-consistent field and the corresponding fluctuation solution are shown for a plate with two elliptical holes. 

In this boundary-aware framework, the Newton-Raphson scheme can be written as
\begin{equation}
    \ten{K}_{ii}(\vec{\tilde{u}_\text{cur}};\Vec{p}) \Delta {\Vec{\tilde{u}}}_i = - \Vec{g}_i (\vec{\tilde{u}_\text{cur}};\Vec{p})\,,\quad \text{where} \quad \vec{\tilde{u}_\text{cur}}=[\vec{\tilde{u}}_{i,\text{cur}}^T,\vec{0}^T]^T\,,
\end{equation}
and initialised with $\vec{\tilde{u}_\text{cur}}\gets \vec{0}$. We obtain a problem with homogenous Dirichlet boundary conditions, as desired. All MOR techniques outlined below will be applied in tandem with a boundary handling scheme along these lines. Of course, many other boundary handling strategies are viable with displacement-space MOR. For example, reduction can be applied to free degrees of freedom only, while values at the boundary degrees of freedom are enforced. In strain-space, this is not feasible since there are no boundary strain values which are explicitly known. Throughout this work, we use the lifting approach both in strain and displacement space, for the sake of consistency. When using displacement-space techniques, it makes very little difference whether we apply a lifted approach or an alternative in terms of accuracy and runtime. In our numerical experiments, the lifted approach has been marginally more efficient than its alternatives because it contributed to improved initial guesses and reduced the number of iterations required until convergence from time to time. Please take note that it is not necessary to handle BCs via a fluctuation field in the full FE model in order to be able to do so in a reduced model; training data can be obtained using a solver of choice.

\section{Model order reduction in displacement space}\label{s:PODECSW}

FE simulation models discretise the unknown field (in this case, the displacement fluctuation) into a vector of $D$ unknown variables $\vec{\tilde{u}}\in \mathbb{R}^D$. Even for moderately sized problems with, e.g. $D < 100,000$ degrees of freedom and $|E|<100,000$ elements, solving for these unknowns using the Newton-Raphson scheme in Eq.~\eqref{eq:newton} incurs considerable computational costs. In every iteration, the assembly of $\ten{K}$ and $\vec{g}$ requires calls to the material routine at $|G|$ Gauss points and evaluations of Eqs.~\eqref{eq:elemres} and~\eqref{eq:K} in $|E|$ elements~\cite{DanCasAkk:2020:mor,DanCasAkk:2022:pca,SchFau:2025:MLA}.
Additionally, the sparse, but large stiffness matrix $\ten{K}\in\mathbb{R}^{D\times D}$ needs to be factorised and Eq.~\eqref{eq:newton} solved. While the solver costs of roughly around $\mathcal{O}(D^2)$ dominate as $D\rightarrow\infty$, the time expended on the assembly, which scales with $\mathcal{O}(|G|)$, can not be discounted, and may actually dominate moderately-sized problems~\cite{SchFau:2025:MLA}.

In the parametric setting, the vast search space can, however, be reduced significantly. Solutions for all parameters $\vec{p}\in R^\delta$ lie on a $\delta$-dimensional solution manifold    
$\mathcal{M}_{\vec{\Tilde{u}}}=\{\vec{\Tilde{u}}_i\mid \exists \vec{p}: \vec{g}(\vec{\tilde{u}};\vec{p})=\vec{0} \}, \vec{p} \in \mathbb{R}^\delta$~\cite{DanCasAkk:2020:mor,DanCasAkk:2022:pca,SchFau:2025:MLA}. Projection-based MOR schemes aim to restrict the search for solutions to an approximation space, which in the ideal case is a tight superspace of the solution manifold. Of course, the solution manifold is not known a priori. Instead, snapshot solutions obtained from the original high-fidelity model
\begin{equation}
    \ten{\tilde{U}} = [\vec{\tilde{u}}_1,\vec{\tilde{u}}_2,...\vec{\tilde{u}}_s] \in \mathbb{R}^{D\times s}\,,
\end{equation}
might be available at parameter samples~\cite{BenGugWil:2015:spm}
\begin{equation}
    \ten{p}^s = [\vec{p}_1,\vec{p}_2,...\vec{p}_s] \in \mathbb{R}^{\delta\times s}\,.
\end{equation}
A conceptual illustration of snapshots on a $\delta=1$-dimensional approximation space in a $D=3$-dimensional solution space can be found in Fig.~\ref{fig:POD}.
Before we go on to discuss how we apply strain-space MOR techniques in this parametric setting, we briefly recall the POD-ECSW as a popular displacement-space approach to accelerating simulations of parametric problems. We choose POD-ECSW as a benchmark method because of its robustness and structure-preservation properties~\cite{FarAveCha:2014:drn}. Alternatives include approximate-then-project schemes such as the Discrete Empirical Interpolation Method~\cite{ChaSor:2010:nmr} and the Gappy POD~\cite{EveSir:1995:klp,Wil:2006:ufls}, and Petrov-Galerkin schemes relying on collocation~\cite{CarBarAnt:2017:glp,CheJiNar:2021:lrc}. We do not explore these here because we have encountered robustness issues with these in our previous work~\cite{FauSch:2025:HML}, and because ECSW has compared favourably to them in our preliminary investigations.

\subsection{Displacement-space POD}

\begin{figure}[h]
    \centering
    \includegraphics[width=0.5\linewidth,trim={2cm 8cm 2cm 6cm},clip]{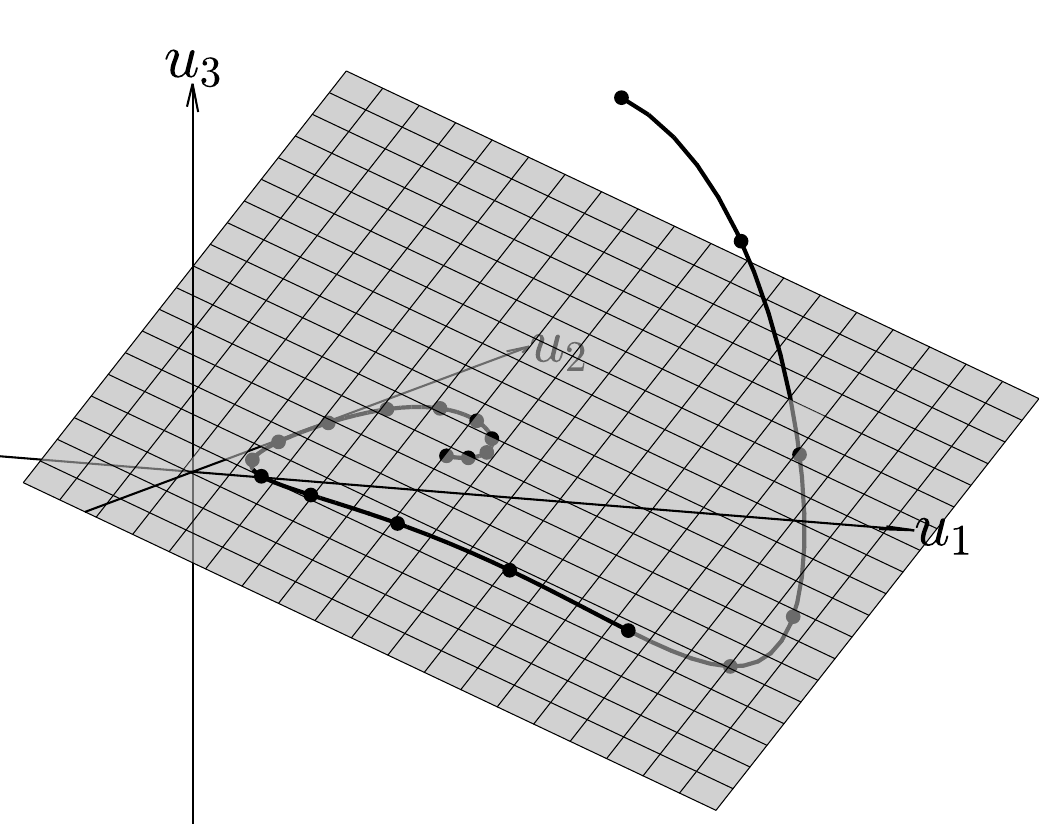}
    \caption{Illustration of a $\delta=1$-dimensional solution manifold (black line) in a $D=3$-dimensional solution space. A $d=2$-dimensional POD model (gray plane) obtained from snapshots (black dots) is also shown. Figure by the authors, originally used in~\cite{FauSch:2026:EMS}.}
    \label{fig:POD}
\end{figure}

The POD introduces a linear approximation of the unknown variables
\begin{equation}
    \vec{\tilde{u}} = \vec{\psi} \vec{y}\,,\label{eq:POD}
\end{equation}
in terms of $d$ reduced unknown variables $\vec{y}\in \mathbb{R}^d$.
A least-squares-optimal mode matrix $\ten{\psi}\in \mathbb{R}^{D\times d}$ can be computed via a truncated thin singular value decomposition (SVD) of the snapshot data $\ten{U} = \ten{\psi} \ten{\Sigma} \ten{R}$~\cite{Pea:1901:llp,Wei:2019:tpo}.
With $\delta \vec{\tilde{u}} = \vec{\psi} \delta \vec{y}$, the discretised weak form in Eq.~\eqref{eq:weakform} becomes
\begin{equation}
    \delta \vec{y}^T \ten{\psi}^T \vec{g} = \delta \vec{y}^T \vec{g}^{\text{red}} = \vec{0}\,,
\end{equation}
and the stiffness matrix is given by
\begin{equation}
    \ten{K}^{\text{red}} = \frac{\partial \vec{g}^{\text{red}}}{\partial \vec{y}} = \ten{\psi}^T \frac{\partial \vec{g}}{\partial \vec{\tilde{u}}} \frac{\partial \vec{\tilde{u}}}{\partial \vec{y}} = \ten{\psi}^T \ten{K}  \ten{\psi}\,,
\end{equation}
such that the Newton-Raphson scheme, when projected onto the reduced space, can be written as
\begin{equation}
    \ten{K}^{\text{red}} \Delta \vec{y} = -\vec{g}^{\text{red}}\,.
\end{equation}
The search for solutions is therefore confined to a $d$-dimensional linear subspace of the original solution space $\mathbb{R}^D$, which cuts the cost of solving linear equation systems significantly~\cite{RadRee:2014:mre}. Finally, the whole displacement solution can be reconstructed from the fluctuation via
\begin{equation}
    \vec{u} = \ten{\psi} \vec{y} + \sum_{j=1}^B \vec{\bar{u}}_j d_j = \vec{\psi} \vec{y}  + \ten{\bar{U}} \vec{d}\,.
\end{equation}
Fig.~\ref{fig:POD} also features a $d=2$-dimensional POD model of the $\delta=1$-dimensional solution manifold.

Because we apply the POD only to the fluctuation field $\vec{\tilde{u}}$, minimising the variance to be captured in this field promotes MOR accuracy. As discussed in Section~\ref{s:bc}, we therefore aim to choose reference parameters to compute $\vec{\bar{u}}_j$ which produce solutions that are close to those expected over the parameter range $p\subset \mathbb{R}^\delta$. Here, we simply choose these parameters to lie in the centre of the parameter domain to be sampled. Because this appears to work well for the examples explored here, we do not investigate alternative strategies to compute $\vec{\bar{u}}_j$. Should a need for higher levels of accuracy arise, alternatives can be explored in future work.

\subsection{Displacement-space POD-ECSW}

The assembly of the reduced residual can conveniently be written on the element level as 
\begin{equation}
    \vec{g}^{\text{red}} = \ten{\psi}^T \sum_{e\in E} \ten{S}^e \vec{g}^e  = \sum_{e\in E} \ten{\psi}^T \ten{S}^e \vec{g}^e  = \sum_{e\in E} \ten{\psi}^{eT} \vec{g}^e\,,
\end{equation}
where $\ten{\psi}^e$ denotes the rows of the mode matrix $\ten{\psi}$ which correspond to the degrees of freedom at the nodes contained in element $e$. While the Galerkin projection thus cuts memory costs significantly, the evaluation of the material routine at all Gauss points still incurs runtime costs which scale linearly with the number of Gauss points $|G|$. The popular ECSW hyperreduction technique cuts these costs by assembling only $|E_\text{m}|<|E|$ elements in a reduced integration domain~\cite{AnKimDou:2008:OCE,FarAveCha:2014:drn}. Alternatively, the assembly can be performed on the Gauss point level using popular ECM technique~\cite{HerCaiFer:2017:dhn,Her:2020:MMP}. The approach used in this work is based on the descriptions in~\cite{FarAveCha:2014:drn}. 

The residual is approximated as a sum over the reduced element set $E_\text{m}$ 
\begin{equation}
    \vec{g}^{\text{red}}  = \sum_{e\in E_\text{m}} \xi^e \ten{\psi}^{eT} \vec{g}^e\,,\label{eq:ECSW_g}
\end{equation}
with positive, volume scaling coefficients $\xi^e$ for each selected element.
The set of reduced elements $E_\text{m}$ and coefficients $\vec{\xi}$ are identified from snapshot data, meaning that element residual snapshots $\vec{g}_j^e, j=1,..,s_g, e \in E$ need to be collected. To enforce consistency, we project the snapshots $\ten{\tilde{U}}$
for parameters $\ten{p}^s$ onto the POD approximation space and gather the element residuals corresponding to these. Alternatively, reduced simulations using the POD can be conducted and element residuals obtained from these. ECSW can be trained with residuals obtained from either converged solutions or unconverged iterates. In this work, we use converged solutions only, in which case it is critical to train not with the entire residual (which vanishes at all snapshots), but only the internal reaction to external loads and and boundary conditions\footnote{An alternative strategy is to impose an ECM-like volume constraint, as discussed in Subsection~\ref{s:ECM}.}.

$E_\text{m}$ and $\vec{\xi}$ can be identified using the sparse nonnegative least squares (NNLS) problem
\begin{equation}
    \min_{\xi_e, E_\text{m}} \sum_j \| ( \sum_{e\in E} \ten{\psi}_j^{eT}\vec{g}_j^e-\sum_{e\in E_\text{m}}  \xi^e \ten{\psi_j}^{eT}\vec{g}_j^e) \|\,,\label{eq:NNLS_ECSW}
\end{equation}
which enforces that the virtual (internal) work done by the reduced model equals the virtual work done by the full model at the snapshots. 
Eq.~\eqref{eq:NNLS_ECSW} can be solved using a greedy Lawson-Hanson procedure until a specified tolerance or until a fixed number $|E_\text{m}|$ of nonzero coefficients $ \xi^e$ have been identified~\cite{FarAveCha:2014:drn}. The latter approach is used here. The elements corresponding to the identified nonzero coefficients define the reduced integration domain $E_\text{m}$. The hyperreduced residual can finally be approximated via Eq.~\eqref{eq:ECSW_g} at a computational cost which scales roughly linearly with $|E_\text{m}|$. An example element set $E_\text{m}$ selected for the simpler of the two examples investigated in this work is shown in Fig.~\ref{fig:ECSW_ECM_RID}.

The stiffness matrix corresponding to Eq.~\eqref{eq:ECSW_g} can be assembled at similar computational cost via
\begin{equation}
    \ten{K}_{\text{red}} = \frac{\partial \vec{g}_{\text{red}}}{\partial \vec{y}} = \sum_{e\in E_\text{m}} \xi^e \ten{\psi}^{eT} \frac{\partial \vec{g}^e}{\partial \vec{y}} = \sum_{e\in E_\text{m}} \xi^e \ten{\psi}^{eT} \frac{\partial \vec{g}^e}{\partial \vec{u}^e} \frac{\partial \vec{u}^e}{\partial \vec{y}} = \sum_{e\in E_\text{m}} \xi^e \ten{\psi}^{eT} \ten{K}^e \ten{\psi}^e\,.
\end{equation}
For an memory- and runtime-efficient implementation of ECSW, it is important to only compute and store the degrees of freedom of $\ten{\psi}$ and $\vec{\tilde{u}}$ which are required for the reduced assembly procedure. 

\section{Model order reduction in strain space}\label{s:strain_MOR}

The displacement-space formulation of POD-ECSW outlined in the previous section necessitates operations on the element level, as well as evaluations of the material routine at all Gauss points within one element~\cite{FarAveCha:2014:drn}. ECM instead performs reduced integration on the Gauss point level, allowing for a more efficient allocation of queries to the material routine~\cite{HerCaiFer:2017:dhn,Her:2020:MMP}. In the context of simulations on RVEs, E3C furthermore conducts the Galerkin-reduction directly in strain space and bypasses displacements (and elements) entirely~\cite{WulHau:2025:ech,wul:2025:ECC,WulHau:2025:EHR}. E3C and a recent variant of ECM with varying weights also circumvent the constraint that the reduced integration points need to be a subset of the original Gauss points, allowing for more efficient integration with fewer integration points~\cite{WulHau:2025:ech,wul:2025:ECC,WulHau:2025:EHR,HerBraPar:2024:CCE}. In this Section, we combine these ideas and allow strain-space MOR techniques to be applied to problems with nonzero Dirichlet BCs using the boundary handling methodology discussed in Section~\ref{s:bcnew}. We pose a version of ECM in strain space and generalise E3C to arbitrary mechanical problems (not just RVEs with periodic BCs).

\subsection{Strain-space POD}

Rather than approximating nodal displacement degrees of freedom, we can use a Galerkin Ansatz for strain measures (see e.g.~\cite{WulHau:2025:ech,FauSch:2026:EMS}), such as for the displacement gradient fluctuation $\vec{\tilde{\underline{H}}}^g$ at the integration points $g \in G$, i.e. 
\begin{equation}
    \vec{\tilde{\underline{H}}}^g = \ten{\varphi}^g \vec{y}\,, g\in G\,.\label{eq:PODH}
\end{equation}
Here and in the following, the underline $\vec{\underline{\cdot}}$ indicates that the tensor quantity $\vec{\cdot}$ has been rewritten in Voigt notation. $\ten{\varphi}^g\in \mathbb{R}^{9\times d}$ refers to the mode matrix corresponding to the entries of the displacement gradient fluctuation at Gauss point $g$.
As in the displacement-space approach discussed above, we state the problem in terms of a fluctuation from a boundary-consistent field
\begin{equation}
    \vec{\tilde{\underline{H}}}^g 
    =\vec{\underline{H}}^g - \sum_j d_j \vec{\bar{\underline{H}}}^g_j\,,
\end{equation}
which is simply the reference-configuration gradient of its displacement-space counterpart in Eq.~\eqref{eq:fluc}. Here, it becomes clear why we can not simply construct our boundary-consistent displacement field by setting the values on the boundary to unity and the rest of the field to zero. This would lead to extreme values and jumps in the BC-consistent strain field close to the boundary, which would have to be countered by the fluctuation field. As a result, basis quality and accuracy would suffer. We can again rewrite the sum as a matrix-vector multiplication
\begin{equation}
    \vec{\tilde{\underline{H}}}^g = \vec{\underline{H}}^g - \ten{\bar{\underline{H}}}^g \vec{d}\,,
\end{equation}
with $\ten{\bar{\underline{H}}}= [\vec{\bar{\underline{H}}}_j]_{j\in [1,B]}$, $\vec{d}= [d_j]_{j\in [1,B]}$, such that the deformation gradient can be computed as 
\begin{equation}
    \vec{\underline{F}}^g = \vec{\tilde{\underline{H}}}^g + \ten{\bar{\underline{H}}}^g \vec{d} + \ten{\underline{I}}\,.
\end{equation}
Here, $\ten{\bar{\underline{H}}}^g$ can be obtained from Gauss point strain values corresponding to the BC-consistent displacement fields~$\ten{\bar{U}}$.

The displacement gradient fluctuation modes $\ten{\varphi}$ can either be computed via an SVD of Gauss point strain values, or derived (by multiplication with the FE strain-displacement matrices) from displacement-space POD modes. Applying an SVD directly to the difference between the displacement gradient field and its boundary-consistent component is numerically advantageous, as it results in orthonormal modes which optimally represent the strain-space snapshots. For the strain-space solvers outlined in the following, this generally leads to slightly higher levels of accuracy than strain-space modes computed from displacement-space modes. The difference, however, is not large, and both strategies are viable. 

Searching for a BC-consistent displacement gradient fluctuation field makes all Dirichlet BCs homogeneous. Because the modes computed from the fluctuation field also vanish at these boundaries, Dirichlet BCs are fulfilled by construction. Furthermore, strain compatibility is retained, because the snapshot data lives in the linear subspace in which compatibility is fulfilled.

The POD-Galerkin Ansatz for the displacement gradient field in Eq.~\eqref{eq:PODH} implies $\delta \vec{\tilde{\underline{F}}}^g =\delta \vec{\tilde{\underline{H}}}^g = \ten{\varphi}^g \delta \vec{y}$.
Substitution into the discretised weak form yields
\begin{equation}
    \delta \vec{y}^T \sum_{g\in G} \ten{\varphi}^{gT} \vec{\underline{P}}(\vec{\underline{F}}^g) V^g = 0\,, \quad \text{such that} \quad \vec{g}^{\text{red}}  =\sum_{g\in G} \ten{\varphi}^{gT} \vec{\underline{P}}(\vec{\underline{F}}^g) V^g\,, \label{eq:gredH}
\end{equation}
and the reduced stiffness matrix becomes
\begin{equation}
    \ten{K}^{\text{red}} = \frac{\partial \vec{g}^{\text{red}}}{\partial\vec{y}} = \sum_{g\in G} \ten{\varphi}^{gT} \frac{\partial \vec{\underline{P}}(\vec{\underline{F}}^g)}{\partial \vec{\underline{F}}^g} \frac{\partial \vec{\underline{F}}^g}{\partial \vec{y}} V^g = \sum_{g\in G} \ten{\varphi}^{gT} \ten{\mathbb{\underline{A}}}(\vec{\underline{F}}^g) \ten{\varphi}^g V^g\,.\label{eq:KredH}
\end{equation}
Eqs.~\eqref{eq:gredH} and.~\eqref{eq:KredH} highlight one of the main benefits of strain-space MOR: element-level operations are no longer necessary, and the projection operation is simplified significantly.

Finally, we require a reconstruction of the displacement fluctuation field 
\begin{equation}
    \vec{\tilde{u}} = \ten{\chi} \vec{y}\,,\label{eq:recon}
\end{equation}
for postprocessing purposes. If the strain-space modes $\ten{\varphi}$ have been derived from displacement-space modes $\ten{\psi}$, the reconstruction matrix $\ten{\chi}$ is simply the displacement-space POD mode matrix from Eq.~\eqref{eq:POD}, i.e.
\begin{equation}
    \ten{\chi} = \ten{\psi}\,.
\end{equation}
In this case, reproduced displacements and strains are mutually compatible by construction.
If $\ten{\varphi}$ was computed directly using an SVD of the displacement gradient snapshots, on the other hand, we can simply use a least squares reconstruction
\begin{equation}
    \ten{\chi} = \ten{\tilde{U}} \ten{Y}^T (\ten{Y} \ten{Y}^T)^{-1}\,,
\end{equation}
where $\ten{Y}=[\vec{y}_i]_{i\in s}\in \mathbb{R}^{d\times s}$. The columns of $\ten{\chi}$ can then be interpreted as (non-orthogonal) displacement fluctuation modes corresponding to the embedding $\ten{Y}$, which was obtained from the strain-space POD. This latter approach has the advantage of allowing us to use proper orthogonal modes in our strain-space solver, with the aforementioned benefits concerning numerical accuracy. While the integration-point strains retain compatibility by construction, the reconstructed displacement field is then no longer exactly consistent with these strains.
In either case, the boundary lifting in Eq.~\eqref{eq:fluc} is applied to obtain the full displacements from the fluctuation. 

We have implemented and tested both approaches on the first of the examples outlined below. Results are nearly identical, but the variant performing the POD directly in strain space tends to be marginally more accurate and computationally efficient. We choose empirical performance over elegance here, and employ it in the investigations below.
More advanced schemes, e.g. employing manifold learning techniques~\cite{SchFau:2025:MLA}, could also be employed for a more accurate reconstruction, but this is beyond the scope of this work.

\subsection{Strain-space POD-ECM}\label{s:ECM}

In~\cite{FauSch:2026:EMS}, we introduced a strain-space variant of ECM for computational homogenisation problems. This variant chooses a subset of Gauss points as strain-space integration points $c\in H \subset G$ 
\begin{equation}
    \vec{g}^{\text{red}} =  \sum_{c\in H} \ten{\varphi}^{cT} \vec{\underline{P}}(\vec{\underline{F}}^c) \xi^c = 0\,,\label{eq:ecm}
\end{equation}
and enlarges the corresponding effective volumes $\xi^c$.
We defined an ECSW-like sparse NNLS problem to identify the set of reduced integration points $H$ and weights $\xi^c$
\begin{equation}
    \min_{\xi^c,H} \sum_{j=1}^s \| ( \sum_{g\in G} \ten{\varphi}^{gT}\vec{\underline{P}}(\vec{\underline{F}}^{gj}) V^g-\sum_{c\in H} \xi^c \ten{\varphi}^{cT} \vec{\underline{P}}(\vec{\underline{F}}^{gj})) \|^2 \quad \text{s.t.} \quad H\subset G, \quad |H|=m\,,
\end{equation}
while also weakly imposing a volume constraint using a penalty term
\begin{equation}
    \sum_{c\in H} \xi^c = \sum_{g \in G} V^g\,,\label{eq:volconst}
\end{equation}
as proposed by~\cite{HerCaiFer:2017:dhn,Her:2020:MMP}. When using full residuals for training, this constraint is necessary to make the sparse NNLS problem well-posed. Otherwise, a trivial solution (with vanishing weights) can exist, because the training residual snapshots are zero to a numerical tolerance if they correspond to converged solutions. Alternatively, we can train only for the internal force part of the residual, or use not the fluctuation modes but the full displacement gradient modes in the NNLS problem. Either approach also makes the problem well-posed. We chose the volume constraint here, because we observe that this can contribute to the robustness of the reduced models\footnote{If stress in the range of the stresses experienced by the structure are multiplied by volumes which add up to the total volume of the structure, the force response of the reduced model is unlikely likely to land in a totally unrealistic range.}.
It is worth noting that, unlike in ECSW, only stress and strain data are required for the weight computation; residual snapshots are bypassed.

Finally, the reduced stiffness matrix follows from Eq.~\eqref{eq:ecm} as
\begin{equation}
    \ten{K}^{\text{red}}= \sum_{c\in H} \ten{\varphi}^{cT} \ten{\mathbb{\underline{A}}}(\vec{\underline{F}}^c) \ten{\varphi}^c \xi^c\,.
\end{equation}
Both this stiffness matrix and the reduced residual in Eq.~\eqref{eq:ecm} operate directly on the Gauss point level and bypass element formulations (for reference, consider the displacement-space ECSW equivalent in Eq.~\eqref{eq:ECSW_g}). This makes the strain-space variant of ECM more efficient and simplifies the implementation considerably. An example set $H$ of reduced integration points selected for the simpler of the two examples investigated in this work is shown in Fig.~\ref{fig:ECSW_ECM_RID}.

\subsection{E3C}\label{s:E3C}

As noted in~\cite{WulHau:2025:ech,wul:2025:ECC,WulHau:2025:EHR,HerBraPar:2024:CCE}, the integration points of a reduced model need not be a subset of the 
integration points of the underlying high-fidelity model.
E3C exploits this observation to search for integration point bases $\ten{\varphi}^c$ at points in strain space which result in accurate reduced integration, but can otherwise be arbitrary~\cite{WulHau:2025:ech,wul:2025:ECC,WulHau:2025:EHR}.

The $\ten{\varphi}^c$ are initialised by grouping all $\ten{\varphi}^g$ into nonoverlapping clusters $C^c\subset G$ using Lloyd's algorithm, and defining centroids
\begin{equation}
    \ten{\varphi}^c = \frac{1}{\xi^c} \sum_{g\in C^c} \ten{\varphi}^g V^g\,,
\end{equation}
with 
\begin{equation}
    \xi^c = \sum_{g\in C^c} V^g\,.
\end{equation}
Then, E3C minimises the snapshot residuals
\begin{align*}
    & \min_{\ten{\varphi}^c} \sum_{j=1}^s \|\sum_{c\in H} \ten{\varphi}^{cT} \vec{\underline{P}}(\vec{\underline{F}}^{cj}) \xi^c \|^2 \,, \quad \\
    & \text{s.t}  \quad \sum_{c\in H} \vec{\tilde{\underline{H}}}^c \xi^c = \sum_{g\in G} \vec{{\tilde{\underline{H}}}}^g V^g\,, \quad \text{where} \quad \vec{{\tilde{\underline{H}}}}^{cj} = \ten{\varphi}^c \vec{y}^j\,,
\end{align*}
subject to a mean fluctuation constraint. While~\cite{WulHau:2025:ech,wul:2025:ECC,WulHau:2025:EHR} utilise a fine-tuned training strategy which generates new training data on demand, we simply train with a fixed snapshot set to facilitate fair comparisons. An L-BFGS algorithm is used for minimisation and run until convergence or stagnation. Since this generally requires a considerable number of iterations, E3C features the most expensive offline phase of the methods explored in this work.

In terms of the online phase, E3C is equivalent to the strain-space variant of ECM discussed in Subsection~\ref{s:ECM}. As the integration points are chosen more freely, however, desirable performance can be achieved with very few integration points.

\begin{figure}
    \centering
    \def\svgwidth{0.9\linewidth}
    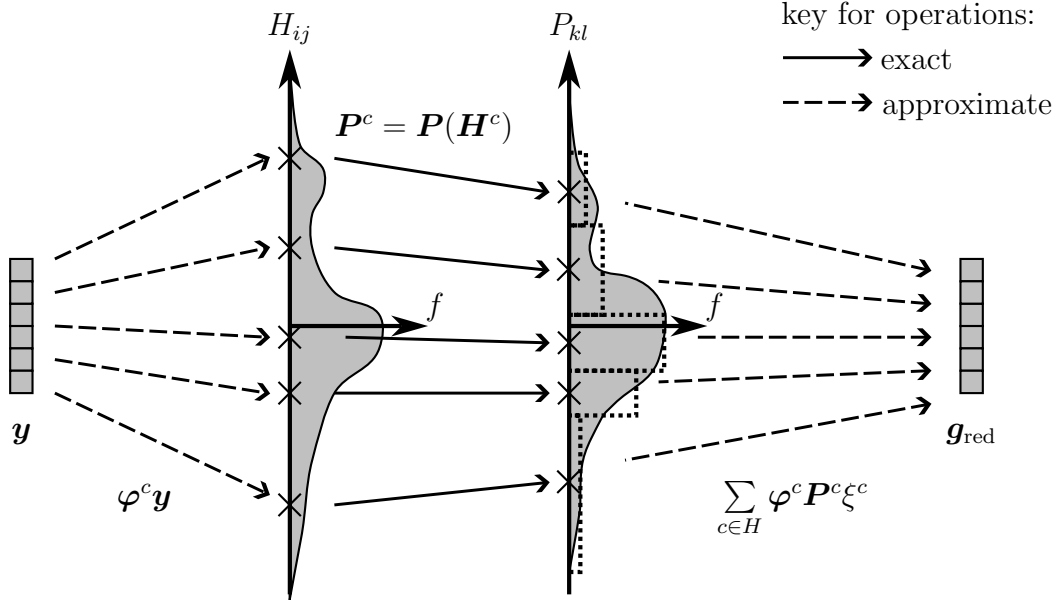
    \caption{Schematic visualisation of reduced cubature: integration point modes allow for strain samples to be inferred from the reduced space. The material law is used to obtain the corresponding stresses. A Gauss-like cubature is performed to approximate the reduced residual. Based on a figure by the authors, in~\cite{FauSch:2026:EMS}.}
    \label{fig:RC}
\end{figure}

As discussed in~\cite{FauSch:2026:EMS}, the approach to hyperreduction taken by E3C and strain-space ECM can be summarised as follows: the strain-space modes at reduced integration points allow us to infer samples of the strain distribution from the reduced variables. At these samples, the material law can be evaluated to obtain the corresponding stresses, which can be projected onto the reduced space using the Galerkin Ansatz. Finally, the multiplication with effective integration point weights and subsequent summation amount to a inexact reduced cubature rule. E3C and ECM might therefore be said to approximately integrate the exact material law when computing the residual. A Newton-Raphson solver is used to iteratively search for solutions, as in a full-order FE scheme. A schematic visualisation of this approach to reduced cubature shown in Fig.~\ref{fig:RC}

\section{Empirical Material Sampling and Linearisation}\label{s:EMSL}

\begin{figure}
    \centering
    \def\svgwidth{0.9\linewidth}
    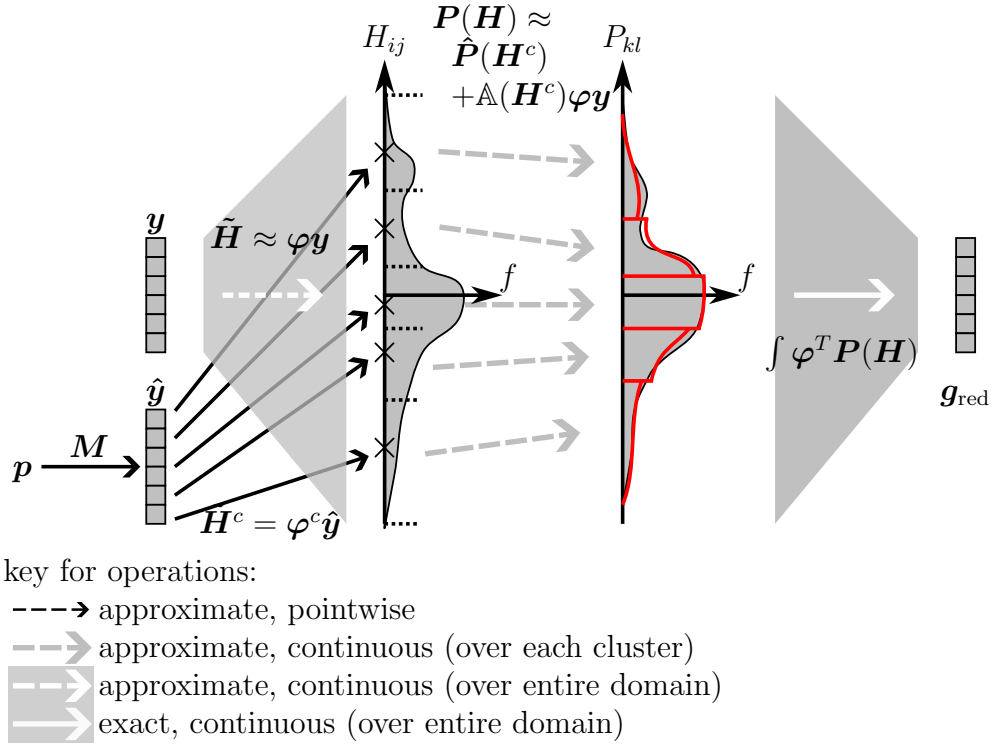
    \caption{Schematic visualisation of EMSL: firstly, an estimate of the reduced solution $\vec{\hat{y}}$ for some parameter value $\vec{p}$ is computed with using linear map $\ten{M}$. Cluster centroid bases $\ten{\psi}^c$ then allows us to estimate the expected average strain in each cluster. The variation of the strain within the cluster, meanwhile, is parameterised using the reduced variables $\vec{y}$, which are yet to be identified. The material behaviour in each cluster is then linearised around the expected cluster average strain, which allows for integration to be performed exactly. Based on a figure by the authors, in~\cite{FauSch:2026:EMS}.}
    \label{fig:EMSL}
\end{figure}

In contrast to ECSW, ECM, and E3C, EMSL does not retain the Newton-Rapshon architecture of the original solver. Instead, the material behaviour is linearised once per load step, around a cluster-wise expected average strain value. This results in an affine problem and allows for exact integration using operations which can mostly be preprocessed in the offline phase. Rather than approximately integrating an exact material law, EMSL thus exactly integrates an approximation of the material law. A schematic visualisation of this approach can be found in Fig.~\ref{fig:EMSL}. In the context of RVEs, we proposed EMSL in~\cite{FauSch:2026:EMS}, and generalise it to problems with arbitrarily-valued Dirichlet BCs here.

EMSL first aims to linearise the material behaviour in a (possibly discontinuous) subdomain of the material around the expected average strain in that subdomain. We therefore need to define appropriate subdomains, as well as a mechanism to estimate the expected strain in these. Subdomains are found by grouping the Gauss points of the underlying high-fidelity model $G$ into nonoverlapping clusters $C^c\subset G$, as in E3C. Example clusters computed for the simpler of the two examples investigated in this work are shown in Fig.~\ref{fig:EMSL_RID}.
The centroids
\begin{equation}
    \ten{\varphi}^c = \frac{1}{\xi^c} \sum_{g\in C^c} \ten{\varphi}^g V^g\,.
\end{equation}
can then be used as bases for the expected average strain in each cluster. 

In each load step, we can estimate the reduced variables from the parameter $p$ using a simple least-squares model with
\begin{equation}
    \ten{M} = \ten{Y}^s \vec{p}^{sT} (\vec{p}^s\vec{p}^{sT})^{-1}\,,
\end{equation}
    allowing for inference
\begin{equation}
    \vec{\hat{y}} = \ten{M} \vec{p}\,.
\end{equation}
The expected average deformation gradient in each cluster can then be reconstructed as
\begin{equation}
    \vec{\hat{\underline{F}}}^c = \ten{\varphi}^c \vec{\hat{y}} + \ten{\bar{\underline{H}}}^c \vec{d} + \text{vec}(\ten{I})\,. \label{eq:EMSLc}
\end{equation}
A more advanced inference strategy, for example employing manifold learning techniques (see e.g.~\cite{SchFau:2025:MLA}) or incorporating information from previous load steps might improve performance and robustness. This is beyond the scope of the present work, which focuses on the generalisation of existing strain-space methods beyond computational homogenisation problems.

Once a prediction of the reduced variables $\vec{\hat{y}}$ and the average cluster strain $\vec{\hat{\underline{F}}}^c$ have been obtained, EMSL linearises the material behaviour in the cluster around $\vec{\hat{\underline{F}}}^c$. The goal is to compute a correction $\vec{y}$ to the predicted solution using the resulting linear model.
In each cluster, we introduce a POD approximation of the displacement gradient fluctuation at all Gauss points 
\begin{equation}
    \vec{\tilde{\underline{H}}}^g = \ten{\varphi}^g \vec{y}\,, g\in G\,,\label{eq:EMSLg}
\end{equation}
which allows us to obtain a linearised model of the stress $\vec{\underline{P}}(\ten{{F}}^g)$ at all Gauss points $g\in C\subset G$ in the cluster
\begin{align}
    \vec{\underline{P}}(\ten{{F}}^g) &= \vec{\underline{P}}(\vec{\hat{\underline{F}}}^c)+\ten{\mathbb{\underline{A}}}(\vec{\hat{\underline{F}}}^c)(\ten{\underline{F}}^g-\vec{\hat{\underline{F}}}^c) + \mathcal{O}(\|\ten{\underline{F}}^g-\vec{\hat{\underline{F}}}^c\|^2)\nonumber\\
    &\approx \vec{\underline{P}}(\vec{\hat{\underline{F}}}^c)+\ten{\mathbb{\underline{A}}}(\vec{\hat{\underline{F}}}^c)(\ten{H}^g-\ten{{H}}^c) \nonumber\\
    & = \vec{\underline{P}}(\vec{\hat{\underline{F}}}^c)-\ten{\mathbb{\underline{A}}}(\vec{\hat{\underline{F}}}^c)\ten{H}^c+ \ten{\mathbb{\underline{A}}}(\vec{\hat{\underline{F}}}^c)(\vec{\tilde{\underline{H}}}^g+ \vec{\bar{\underline{H}}}^g \vec{d})\nonumber\\
    & \approx \vec{\underline{P}}(\vec{\hat{\underline{F}}}^c)-\ten{\mathbb{\underline{A}}}(\vec{\hat{\underline{F}}}^c)\ten{H}^c+ \ten{\mathbb{\underline{A}}}(\vec{\hat{\underline{F}}}^c)\ten{\varphi}^g \vec{y}+ \ten{\mathbb{\underline{A}}}(\vec{\hat{\underline{F}}}^c) \vec{\bar{\underline{H}}}^g \vec{d}\,.\label{eq:EMSLp}
\end{align}
Substitution into the residual yields a low-dimensional linear expression
\begin{align*}
    \vec{g}^{\text{red}} & = \sum_{c\in H} \sum_{g \in C^c} \ten{\varphi}^{gT} \vec{\underline{P}}(\ten{{F}}^g) V^g \nonumber \\
    & = \sum_{c\in H} ( \sum_{g \in C^c} \ten{\varphi}^{gT} V^g (\vec{\underline{P}}(\vec{\hat{\underline{F}}}^c)-\ten{\mathbb{\underline{A}}}(\vec{\hat{\underline{F}}}^c)\vec{\underline{H}}^c)  + \sum_{g \in C^c} \ten{\varphi}^{gT} V^g \ten{\mathbb{\underline{A}}}(\vec{\hat{\underline{F}}}^c)\ten{\varphi}^g \vec{y} + \sum_{g \in C^c} \ten{\varphi}^{gT} V^g \ten{\mathbb{\underline{A}}}(\vec{\hat{\underline{F}}}^c) \ten{\bar{\underline{H}}}^g \vec{d}) \nonumber \\
    & = \vec{r} + \ten{T}\vec{y} + \vec{s}\,.\label{eq:EMSLres}
\end{align*}
While the approach is conceptually simple, the efficient computation of the terms $\vec{r}$, $\ten{T}$, and $\vec{s}$ necessitates some rearrangements. In particular, 
\begin{equation}
    \vec{r} = \sum_{c\in H} \ten{\Bar{\varphi}}^c \ten{\hat{\underline{P}}}(\ten{F}^c)\,,
\end{equation}
with
\begin{equation}
    \ten{\Bar{\varphi}}^c = \sum_{g\in C^c} \ten{\varphi}^g V^g\,,
\end{equation}
and
\begin{equation}
    \ten{\hat{\underline{P}}}(\ten{F}^c) = \vec{\underline{P}}(\vec{\hat{\underline{F}}}^c)-\ten{\mathbb{\underline{A}}}(\vec{\hat{\underline{F}}}^c)\vec{\underline{H}}^c\,.\label{eq:EMSLPhat}
\end{equation}
Additionally, we can rewrite $\vec{s}$ and $\ten{T}$ in index notation. Firstly,
\begin{equation}
    s_\alpha = \sum_{c\in H} \mathbb{\underline{A}}_{\beta\gamma}(\vec{\underline{F}}^c)\mathcal{W}_{\beta\alpha\gamma\delta}^c d_\delta\,,
\end{equation}
with
\begin{equation}
    \mathcal{W}_{\beta\alpha\gamma\delta}^c = \sum_{g\in C^c} \varphi_{\beta\alpha}^g \bar{\underline{H}}_{\gamma\delta}^g V^g\,.
\end{equation}
Secondly,
\begin{equation}
    T_{\alpha\beta} = \sum_{c\in H} \sum_{g\in C^c} \varphi_{\gamma\alpha}^g \mathbb{\underline{A}}_{\gamma\delta}(\vec{\underline{F}}^c) \varphi_{\delta\beta}^g V^g = \sum_{c\in H} \mathbb{\underline{A}}_{\gamma\delta}(\vec{\underline{F}}^c) \sum_{g\in C^c} \varphi_{\gamma\alpha}^g \varphi_{\delta\beta}^g V^g = \sum_{c\in H} \mathbb{\underline{A}}_{\gamma\delta}(\vec{\underline{F}}^c) \mathcal{D}_{\gamma\alpha\beta\delta}^c\,,
\end{equation}
with
\begin{equation}
    \mathcal{D}_{\gamma\alpha\beta\delta}^c = \sum_{g\in C^c} \varphi_{\gamma\alpha}^g \varphi_{\delta\beta}^g V^g\,.
\end{equation}
The high-dimensional operations involved in the computation of $\ten{\Bar{\varphi}}^c$, $\ten{\mathcal{W}}^c$, and $\ten{\mathcal{D}}^c$ can therefore be preprocessed in the offline phase. In each load step, we then need only evaluate $\ten{\hat{\underline{P}}}(\ten{\hat{F}}^c)$ and $\ten{\mathbb{\underline{A}}}(\vec{\hat{\underline{F}}}^c)$ at the expected cluster average, and compute $\vec{r}$, $\vec{s}$, and $\vec{T}$ using low-dimensional products with $\ten{\Bar{\varphi}}^c$, $\ten{\mathcal{W}}^c$, and $\ten{\mathcal{D}}^c$, for each cluster $c$.
$\ten{T}$ becomes the reduced stiffness matrix
\begin{equation}
    \ten{K}^{\text{red}} = \ten{T}\,,
\end{equation}
and the incremental problem in each load step becomes affine. This circumvents the need for Newton iterations in the online phase, meaning that, all else being equal, EMSL requires significantly fewer queries to the material law than ECSW, ECM, or E3C. 


\section{Application: hyperelastic material parameter discovery}\label{s:application}

In this work, we apply the MOR techniques discussed above to two parameterised hyperelastic large-deformation solid-mechanical problems inspired by inverse parameter fitting. In particular, we consider the plate with two elliptical holes shown in Fig.~\ref{fig:pw2h}, which has been used for material model discovery in~\cite{FlaKumDel:2022:DPM,FlaKumDel:2023:ADG,ThaGuoSai:2025:CKC,HolKuh:2026:CNN}. 

We employ the compressible Mooney-Rivlin material law in Eq.~\eqref{eq:MR} to model reversible, large-deformation material behaviour~\cite{Hol:2002:nsm}. In the investigations below, we vary the material parameters in ranges of $\lambda \in [400,10000]$, and $c_1, c_2 \in [100,1000]$, respectively. 

In Subsection~\ref{s:tension}, we first subject the plate to a tensile load case with up to $20\%$ overall strain.
In Subsection~\ref{s:tensionshear}, we then test the validity of the boundary handling strategy for multiple nonzero Dirichlet boundaries. The plate is subjected to a tensile load case with up to $50\%$ overall strain, a shear load case with up to $25\%$ overall strain, and a hybrid load case with up to $35\%$ tensile and $20\%$ shear strain. In addition, the parameters of the employed Mooney-Rivlin model are varied, as in the first example. Because strains are larger and strain localisation is more pronounced, a finer discretisation is also required in this case. The second example therefore features a larger variety of loading scenarios, larger deformations, and higher computational costs. We only discuss the results obtained in the simpler first example briefly but take the opportunity to illustrate the behaviour of the investigated MOR techniques. The data obtained from the the more complicated second benchmark are explored in more detail.

\subsection{First example: tension loading, moderate deformations, low dimensionality}\label{s:tension}

\subsubsection*{Setup}

\begin{figure*}[h]
\begin{subfigure}[t]{.45\textwidth}
  \centering
    \def\svgwidth{0.7\linewidth}
    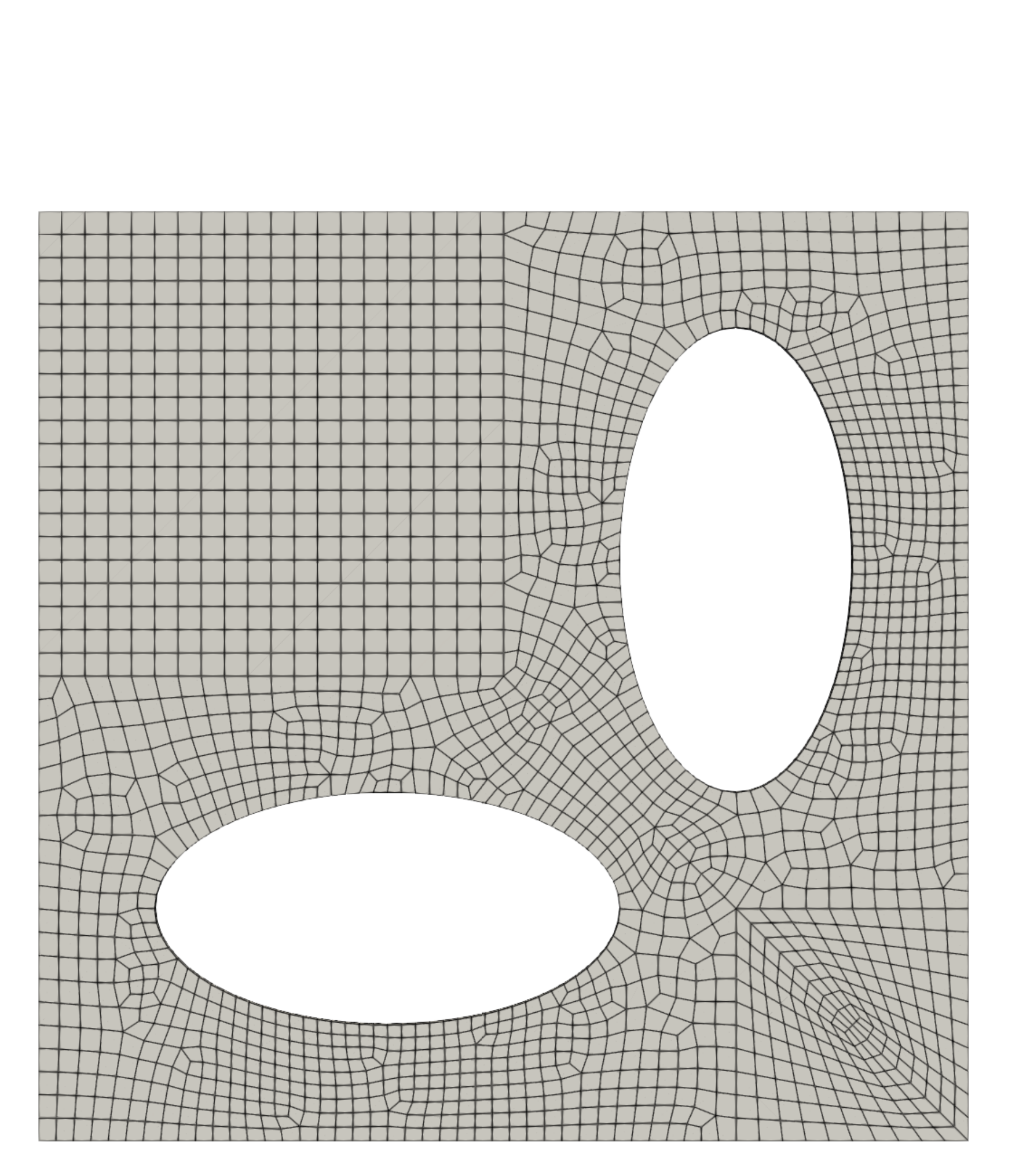
  \caption{Coarse discretisation of plate with two elliptical holes.}
  \label{fig:pw2h1}
\end{subfigure}
\hspace{0.5cm}
\begin{subfigure}[t]{.45\textwidth}
  \centering
    \def\svgwidth{\linewidth}
\begingroup%
  \makeatletter%
  \providecommand\color[2][]{%
    \errmessage{(Inkscape) Color is used for the text in Inkscape, but the package 'color.sty' is not loaded}%
    \renewcommand\color[2][]{}%
  }%
  \providecommand\transparent[1]{%
    \errmessage{(Inkscape) Transparency is used (non-zero) for the text in Inkscape, but the package 'transparent.sty' is not loaded}%
    \renewcommand\transparent[1]{}%
  }%
  \providecommand\rotatebox[2]{#2}%
  \newcommand*\fsize{\dimexpr\f@size pt\relax}%
  \newcommand*\lineheight[1]{\fontsize{\fsize}{#1\fsize}\selectfont}%
  \ifx\svgwidth\undefined%
    \setlength{\unitlength}{1180.15484619bp}%
    \ifx\svgscale\undefined%
      \relax%
    \else%
      \setlength{\unitlength}{\unitlength * \real{\svgscale}}%
    \fi%
  \else%
    \setlength{\unitlength}{\svgwidth}%
  \fi%
  \global\let\svgwidth\undefined%
  \global\let\svgscale\undefined%
  \makeatother%
  \begin{picture}(1,0.82622173)%
    \lineheight{1}%
    \setlength\tabcolsep{0pt}%
    \put(0,0){\includegraphics[width=\unitlength,page=1]{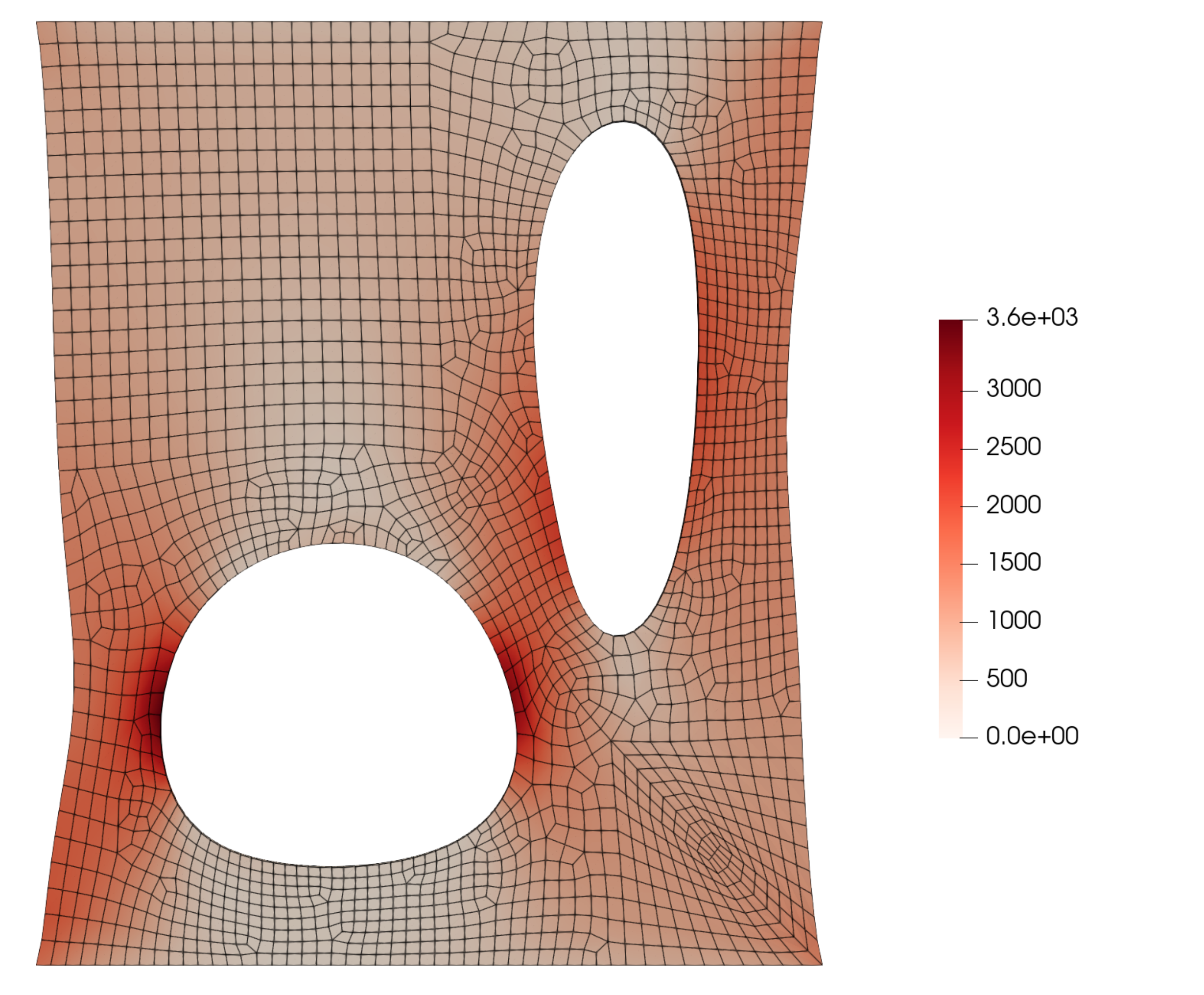}}%
    \put(0.88338088,0.38300359){\color[rgb]{0.01960784,0.01960784,0.01960784}\makebox(0,0)[lt]{\lineheight{1.25}\smash{\begin{tabular}[t]{l}$\sigma^{\text{vM}}$\end{tabular}}}}%
  \end{picture}%
\endgroup%

  \caption{Deformed plate at the end of tensile load path for one parameter sample. Deformation shown to scale, von Mises stress $\sigma^{\text{vM}}$ shown using heatmap.}
  \label{fig:defo1}
\end{subfigure}
\caption{Undeformed configuration of plate with two holes and deformation and stress at the end of tensile load path, for one material parameter sample.}
\label{fig:}
\end{figure*}

For the simpler, tensile benchmark, the plate is discretised using $2022$ linear hexahedral elements and $4382$ nodes, resulting in $13,146$ degrees of freedom. The discretisation is shown in Fig.~\ref{fig:pw2h1}. We are thus considering quite a small problem, meaning that assembly costs are dominant. Repeatedly performing FE simulations for such a component in a context of inverse modeling or fitting nonetheless incurs considerable computational costs, which can be ameliorated using reduced models. We consider an example with higher computational cost in Subsection~\ref{s:tensionshear}. There, a finer discretisation is required due to larger deformations and stronger strain localisation.

In this section, the plate with two holes is subjected to tensile loading via a Dirichlet BC applied to the upper surface, with $d_1 \in [0,20]$, i.e. up to $20\%$ average strain, though local strains exceed $60\%$. $5$ load steps are scheduled for each load path, but the step size is halved if the Newton-Raphson solver diverges. A solution at the end of the load path for example material parameters is shown in Fig.~\ref{fig:defo1}.

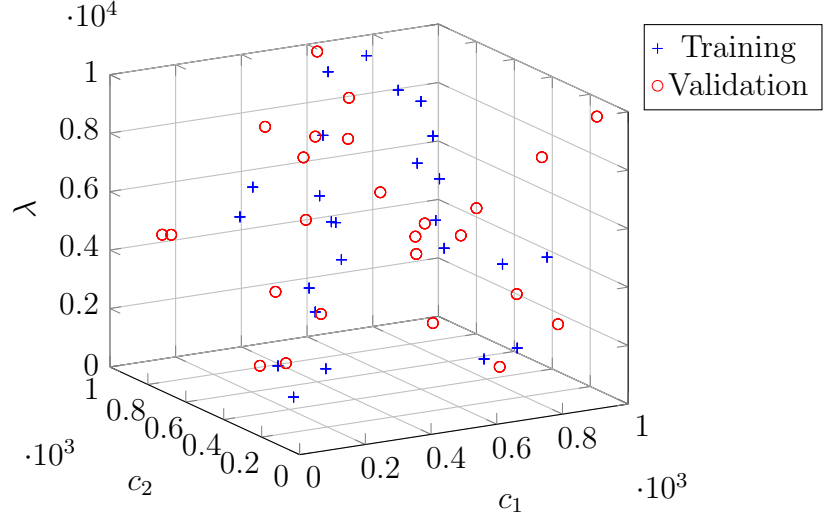
\begin{figure}[h]
\centering
\begin{tikzpicture}
    \begin{axis}
        [view={-30}{60},
        xmin=0,xmax=1000,
        ymin=0,ymax=1000,
        zmin=0,zmax=10000,
        xtick={0,200,400,600,800,1000},
        ytick={0,200,400,600,800,1000},
        ztick={0,2000,4000,6000,8000,10000},
        scaled x ticks=base 10:-3,
        scaled y ticks=base 10:-3,
        xlabel=$c_1$,
        ylabel=$c_2$,
        zlabel=$\lambda$,
        legend pos=outer north east,
        plot box ratio = 1 1 5,
        zmajorgrids=true,
        ymajorgrids=true,
        xmajorgrids=true,
        ]
        
        \addplot3[
        blue,
        only marks,
        mark=+
        ]
        table[z index=0,x index=1,y index=2,col sep=comma] 
        {par_train.csv};
        \addlegendentry{Training}
        
        \addplot3[
        red,
        only marks,
        mark=o
        ]
        table[z index=0,x index=1,y index=2,col sep=comma]
        {par_val.csv};
        \addlegendentry{Validation}
        
    \end{axis}
\end{tikzpicture}
\caption{Training (blue plus signs) and validation (red circles) samples in material parameter space.}
\label{fig:samples}
\end{figure}

$25$ samples of the material parameter values are taken via latin hypercube sampling for both training and for validation, such that across $5$ load steps for each we have $s=125$ snapshots, and $125$ independent validation parameter combinations in the same parameter range. The parameter values for the snapshot and validation sets are both shown in Fig.~\ref{fig:samples}.

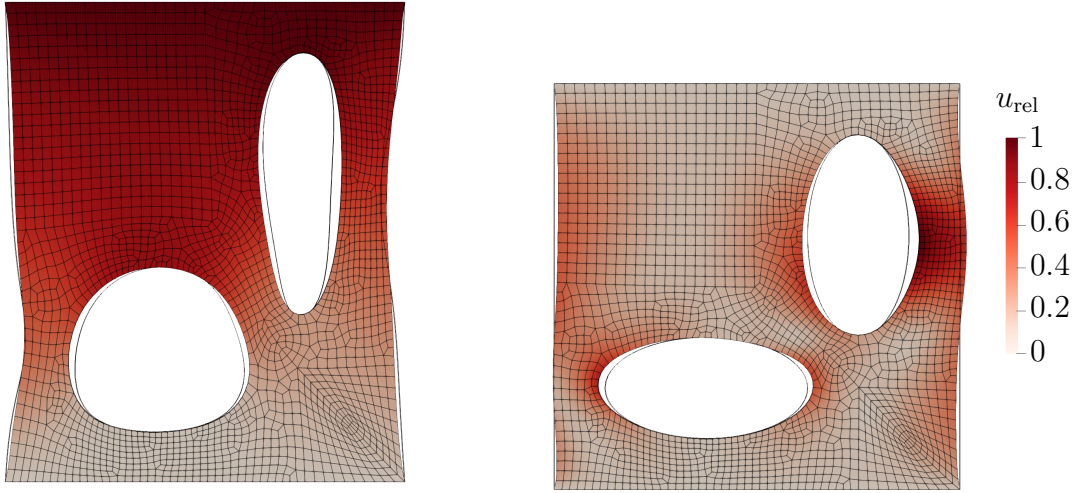
\begin{figure*}[h]
\begin{subfigure}[t]{.45\textwidth}
  \centering
    \def\svgwidth{0.8\linewidth}
    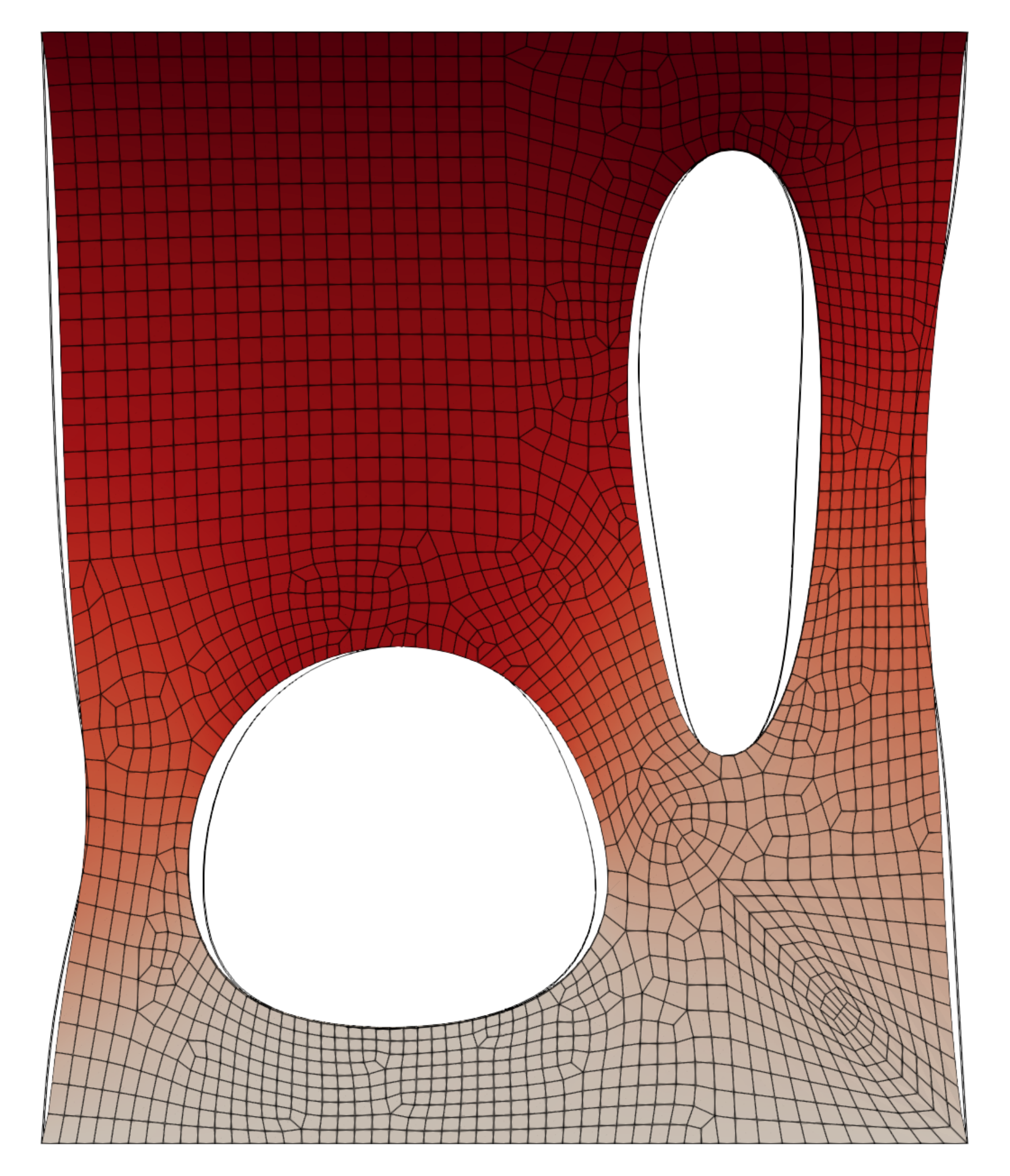
  \caption{Deformation of plate with two holes at the end of tensile load path for one parameter sample, BC-consistent field shown as black outline. Displacement magnitude shown as heatmap.}
  \label{fig:bcfield1}
\end{subfigure}
\hspace{0.5cm}
\begin{subfigure}[t]{.45\textwidth}
  \centering
    \def\svgwidth{1.05\linewidth}
\begingroup%
  \makeatletter%
  \providecommand\color[2][]{%
    \errmessage{(Inkscape) Color is used for the text in Inkscape, but the package 'color.sty' is not loaded}%
    \renewcommand\color[2][]{}%
  }%
  \providecommand\transparent[1]{%
    \errmessage{(Inkscape) Transparency is used (non-zero) for the text in Inkscape, but the package 'transparent.sty' is not loaded}%
    \renewcommand\transparent[1]{}%
  }%
  \providecommand\rotatebox[2]{#2}%
  \newcommand*\fsize{\dimexpr\f@size pt\relax}%
  \newcommand*\lineheight[1]{\fontsize{\fsize}{#1\fsize}\selectfont}%
  \ifx\svgwidth\undefined%
    \setlength{\unitlength}{1087.13946533bp}%
    \ifx\svgscale\undefined%
      \relax%
    \else%
      \setlength{\unitlength}{\unitlength * \real{\svgscale}}%
    \fi%
  \else%
    \setlength{\unitlength}{\svgwidth}%
  \fi%
  \global\let\svgwidth\undefined%
  \global\let\svgscale\undefined%
  \makeatother%
  \begin{picture}(1,0.90002478)%
    \lineheight{1}%
    \setlength\tabcolsep{0pt}%
    \put(0,0){\includegraphics[width=\unitlength,page=1]{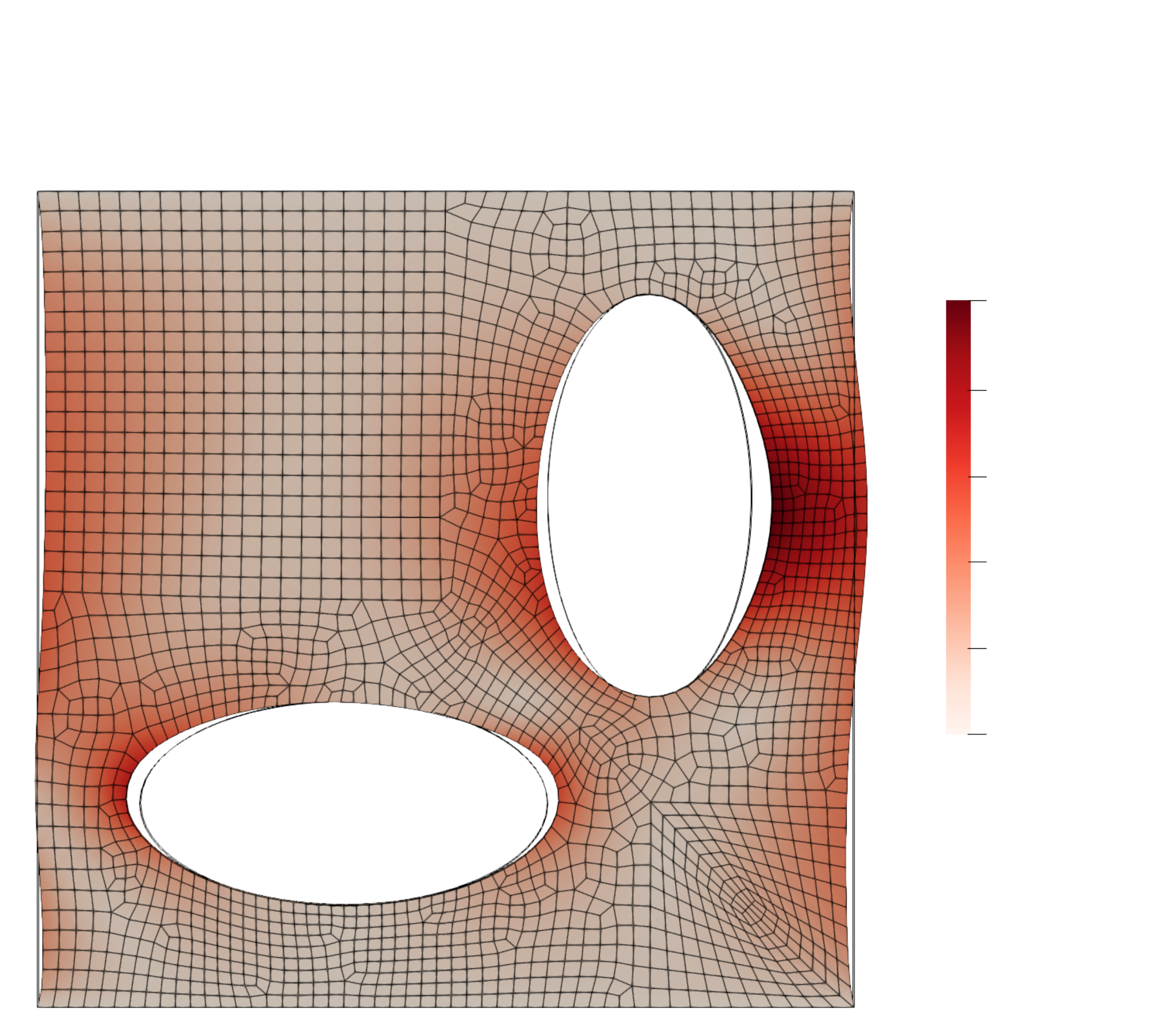}}%
    \put(0.8044221,0.69299215){\color[rgb]{0,0,0}\makebox(0,0)[lt]{\lineheight{1.25}\smash{\begin{tabular}[t]{l}$u_\text{rel}$\end{tabular}}}}%
    \put(0.862881,0.62340927){\color[rgb]{0,0,0}\makebox(0,0)[lt]{\lineheight{1.25}\smash{\begin{tabular}[t]{l}1\end{tabular}}}}%
    \put(0.86451947,0.24834904){\color[rgb]{0,0,0}\makebox(0,0)[lt]{\lineheight{1.25}\smash{\begin{tabular}[t]{l}0\end{tabular}}}}%
    \put(0.86451947,0.32543228){\color[rgb]{0,0,0}\makebox(0,0)[lt]{\lineheight{1.25}\smash{\begin{tabular}[t]{l}0.2\end{tabular}}}}%
    \put(0.86451947,0.39922093){\color[rgb]{0,0,0}\makebox(0,0)[lt]{\lineheight{1.25}\smash{\begin{tabular}[t]{l}0.4\end{tabular}}}}%
    \put(0.86451947,0.47633086){\color[rgb]{0,0,0}\makebox(0,0)[lt]{\lineheight{1.25}\smash{\begin{tabular}[t]{l}0.6\end{tabular}}}}%
    \put(0.86451947,0.55011951){\color[rgb]{0,0,0}\makebox(0,0)[lt]{\lineheight{1.25}\smash{\begin{tabular}[t]{l}0.8\end{tabular}}}}%
  \end{picture}%
\endgroup%

  \caption{Fluctuation field corresponding to solution shown on the left, with reference configuration outlined in black. Displacement fluctuation shown as heatmap.}
  \label{fig:fluc1}
\end{subfigure}
\caption{Deformation at the end of tensile load path, compared to BC-consistent field, and fluctuation from BC-consistent field. Heatmap colouring according to displacement or displacement fluctuation relative to maximum.}
\label{fig:BC}
\end{figure*}

We additionally compute a solution with $\bar{\lambda}$, $\bar{c}_1$, $\bar{c}_2$ chosen in the centre of the sampled parameter range with a deformation of $d_1=1$, which is used as a BC-consistent field. The BC-consistent field (scaled with $d_1$) is shown as a black outline in Fig.~\ref{fig:bcfield1}, where an example solution field is also presented for reference. Fig.~\ref{fig:fluc1} illustrates the corresponding fluctuation field, with the undeformed plate being drawn as a black outline. The variance of the fluctuation solutions is considerably smaller than that of the full displacement field, which simplifies the work of the MOR strategies in accurately predicting this full displacement field.

Once the snapshots and reference solutions have been computed, we also use the MOR techniques discussed in this work in tandem with the boundary handling methodology outlined in Section~\ref{s:bcnew} to compute solutions for the validation material parameters. Then, the mean relative error in the displacement field 
\begin{equation}
    E_\text{mean} = \frac{1}{s} \sum_{i=1}^{s} \frac{\|\Vec{u}^\text{i,ROM}-\vec{u}^\text{i,VAL}\|}{\|\Vec{u}^\text{i,VAL}\|}\,,\label{eq:error}
\end{equation}
is calculated. The runtimes required to complete all validation simulations are also recorded. We subtract the $0.07888$ s overhead to compare only the runtime cost of necessary solver operations.

Absolute runtimes are of course strongly subject to the specifics of the implementation, and must be analysed with care. 
We ran all simulations discussed in this work using our in-house python FE framework on standard laptop hardware with 16 AMD Ryzen 7 PRO 4750U processors and 32 GB RAM, with the same routines being employed wherever possible. The solvers differ only in the selection of integration points and weights, whether the assembly is performed in displacement space on the element level or in strain space on the integration point level, and, in the case of EMSL, in the multiplication of material terms with the necessary preprocessed factors. While the data presented here does not allow for absolute statements about runtime performance, we can tentatively compare relative runtime advantages. 

The runtime required for the high-fidelity model to produce solutions for the entire validation set is $3820$~s (approximately $1$ hour and $3$ minutes) of which $0.05191$ s is overhead incurred due to sampling. In computing these solutions, we utilise the UMPACK multifrontal sparse LU solver, and a compressed-sparse-row assembly procedure with a precomputed sparsity pattern~\cite{dav:2004:a8u}.

To explore the accuracy/runtime tradeoff achievable with a variety of algorithmic parameters, simulations using all MOR techniques are performed for a range of reduced basis sizes $d\in [5,30]$ and reduced integration points $|H|\in[1,100]$. In ECSW, we vary the selected elements $|E_\text{m}|\in[1,100]$, i.e. $|H|\in[6,600]$. Because performance trends over these parameters are analysed in detail for the second example in Subsection~\ref{s:tensionshear}, we only briefly explore these trends here, instead putting an emphasis on illustrating how the different hyperreduction techniques work on an example problem.

\subsubsection*{Results}

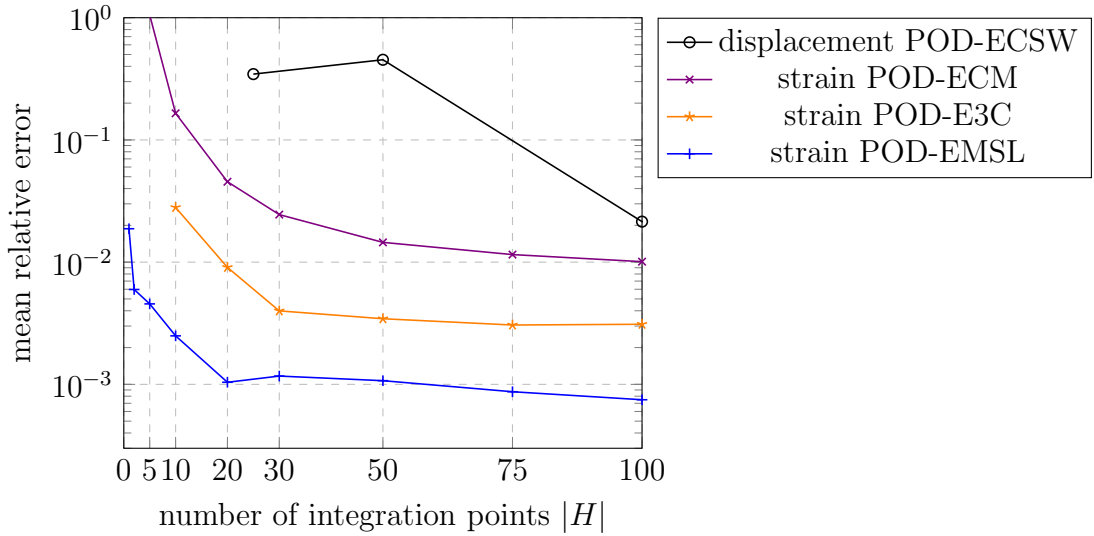
\begin{figure}[h]
    \centering
    \begin{tikzpicture}
    \begin{semilogyaxis}[
        ylabel={mean relative error},
        xlabel={number of integration points $|H|$},
        xmin=0, xmax=100,
        ymin=3e-4, ymax=1e0, domain=3e-4:1e0,
        xtick={0,5,10,20,30,50,75,100},
        legend pos=outer north east,
        ymajorgrids=true,
        xmajorgrids=true,
        grid style=dashed,
        table/col sep=semicolon,
    ]

    \addplot+[
        black,
        semithick,
        mark=o]
    table[x index=5,y index=1] 
    {allemd20.csv};
    \addlegendentry{displacement POD-ECSW}

    \addplot+[
        violet,
        semithick,
        mark=x]
    table[x index=0,y index=2] 
    {allemd20.csv};
    \addlegendentry{strain POD-ECM}

    \addplot+[
        orange,
        semithick,
        mark=star]
    table[x index=0,y index=3] 
    {allemd20.csv};
    \addlegendentry{strain POD-E3C}
    
    \addplot+[
        blue,
        semithick,
        mark=+]
    table[x index=0,y index=4] 
    {allemd20.csv};
    \addlegendentry{strain POD-EMSL}

    \end{semilogyaxis}
    \end{tikzpicture}
    \caption{Mean relative error over number of integration points $|H|$, for $d=20$ and displacement-space ECSW (black circles), as well as strain-space ECM (violet $\times$ symbols), E3C (orange stars), and EMSL (blue plus signs).}
    \label{fig:ed20}
\end{figure}

Fig.~\ref{fig:ed20} shows the mean relative displacement error over a range of integration points $|H|\in [1,100]$, for $d=20$, and all MOR techniques investigated here. To make the comparison fair, the results obtained using ECSW are also plotted over the number of selected integration points $|H|$, not the number of selected elements $|E_\text{m}|$\footnote{The latter is smaller by a factor of $6$, since each linear hexahedral element selected by ECSW contains $6$ Gauss points.}. 

As expected, the error falls with increasing $|H|$ for all methods. In the case of displacement-space ECSW, it drops gradually throughout the investigated range, reaching $1\%$ at $|H|\approx200$. For an error level of below $0.1\%$, $750$ integration points (i.e., $|E_\text{m}|=150$ elements) are required, which is beyond the range shown in this diagram. The strain-space version of ECM, meanwhile, improves on this performance over the whole range of $|H|$, and achieves $1\%$ error with $|H|=100$. Performing the NNLS selection on the integration point level allows ECM to effectively afford $6$ times the number of independent integration weights $\xi^c_c$ for a given $|H|$, which explains this advantage. 

E3C improves upon the performance of ECM by yet another significant step, achieving an error level of below $1\%$ with only $20$ integration points, and below $0.5\%$ with $|H|=30$. The error then drops gradually to $0.3\%$ at $|H|=100$. The more flexible integration point selection, which is no longer constrained to yield a subset of the original integration points, appears to pay off in this metric. 

EMSL also outperforms E3C, yielding errors which are generally half an order of magnitude smaller than those obtained using E3C. Only two reference points are required for errors below $1\%$, and an error level of around $0.1\%$ is achieved using only $20$ such points. In this simple example, exactly integrating the cluster-wise linearisation of the material law seems to yield higher levels of accuracy than the inexact integration of the exact material law favoured by E3C, ECM, and ECSW. A more strongly nonlinear example with larger and more varied deformation states is explored in Section~\ref{s:tensionshear}.

Because we explore trends in the level of accuracy over $d$ and $|H|$ in detail for the second example in Subsection~\ref{s:tensionshear}, we do not discuss these in much detail here. Generally, trends at different $d$ are similar to those seen in Fig.~\ref{fig:ed20}. The reduction in the error levels with increasing $d$ is less pronounced than that with $|H|$. There is generally a drop from $d=5$ to $d=15$, but further increases in $d$ do not make a significant difference, suggesting that the solution manifold lies close to a low-dimensional linear subspace of the solution space.

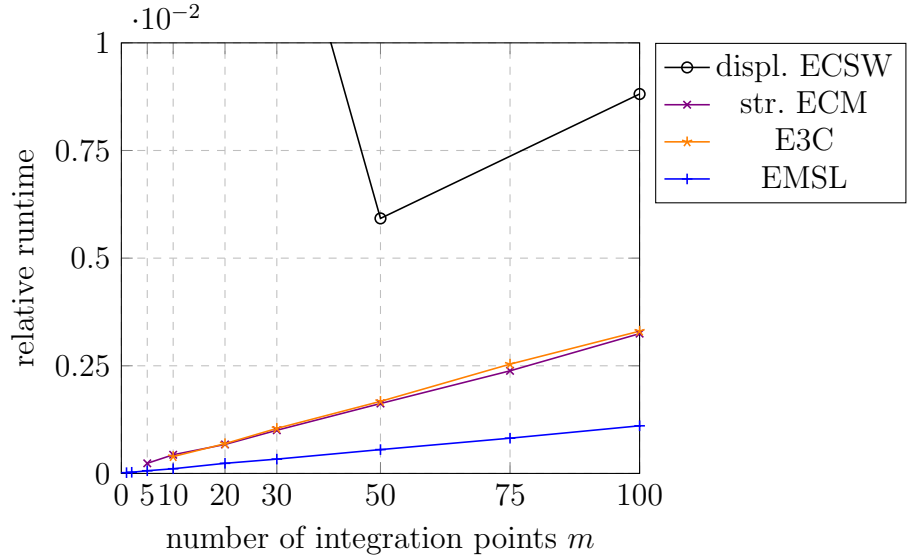
\begin{figure}[h]
    \centering
    \begin{tikzpicture}
    \begin{axis}[
        ylabel={relative runtime},
        xlabel={number of integration points $m$},
        xmin=0, xmax=100,
        ymin=0, ymax=0.01,
        xtick={0,5,10,20,30,50,75,100},
        ytick={0,0.0025,0.005,0.0075,0.01},
        legend pos=outer north east,
        ymajorgrids=true,
        xmajorgrids=true,
        grid style=dashed,
        table/col sep=semicolon,
    ]

    \addplot+[
        black,
        semithick,
        mark=o]
    table[x index=5,y index=1] 
    {alltmd20.csv};
    \addlegendentry{displ. ECSW}

    \addplot+[
        violet,
        semithick,
        mark=x]
    table[x index=0,y index=2] 
    {alltmd20.csv};
    \addlegendentry{str. ECM}

    \addplot+[
        orange,
        semithick,
        mark=star]
    table[x index=0,y index=3] 
    {alltmd20.csv};
    \addlegendentry{E3C}
    
    \addplot+[
        blue,
        semithick,
        mark=+]
    table[x index=0,y index=4] 
    {alltmd20.csv};
    \addlegendentry{EMSL}

    \end{axis}
    \end{tikzpicture}
    \caption{Relative runtime over number of integration points $m$, for $d=10$ for displacement-space (black circles), as well as strain-space ECM (violet $\times$ symbols, E3C (orange stars), and EMSL (blue plus signs).}
    \label{fig:dt20}
\end{figure}

Next, Fig.~\ref{fig:dt20} shows the evolution of the relative runtime over the number of integration points $|H|$, again for $d=20$. As expected, all methods scale close to linearly with $|H|$, as the increase in the number of integration points necessitates proportionately more queries to the material law. Deviations from this trend might occur due to slightly more or fewer Newton-Raphson iterations being required, but these variations do not appear to be significant for ECM and E3C. ECSW encounters an unexpectedly large runtime for $|E|=5$, $|H|=25$ precisely because the solver requires dramatically more iterations as load stepping adaptivity is activated repeatedly. As can be seen from the elevated error level in Fig.~\ref{fig:ed20} corresponding to this parameter combination, the ECSW model does not accurately capture the system behaviour in that instance. Over the remainder of the parameter range, the displacement-space ECSW model requires slightly more than twice as much runtime as a strain-space ECM or E3C model. ECSW tends to require slightly more iterations until convergence, but the runtime deficit is mostly due to the additional cost of element-level operations. The reduced complexity of strain-space MOR techniques thus directly leads to improved runtimes in the online phase. 

Unsurprisingly, given that the solver formulations are identical, ECM and E3C result in almost identical runtimes. For a given hyperreduced model size $|H|$, EMSL then slashes the runtimes required by ECM and E3C by more than half again. The affine formulation means that iterations within a load step are not required, which reduces the number of required queries to the material routine significantly.

\begin{figure}[h]
    \centering
    \includegraphics[width=0.5\linewidth,trim={15cm 3.5cm 15cm 5cm},clip]{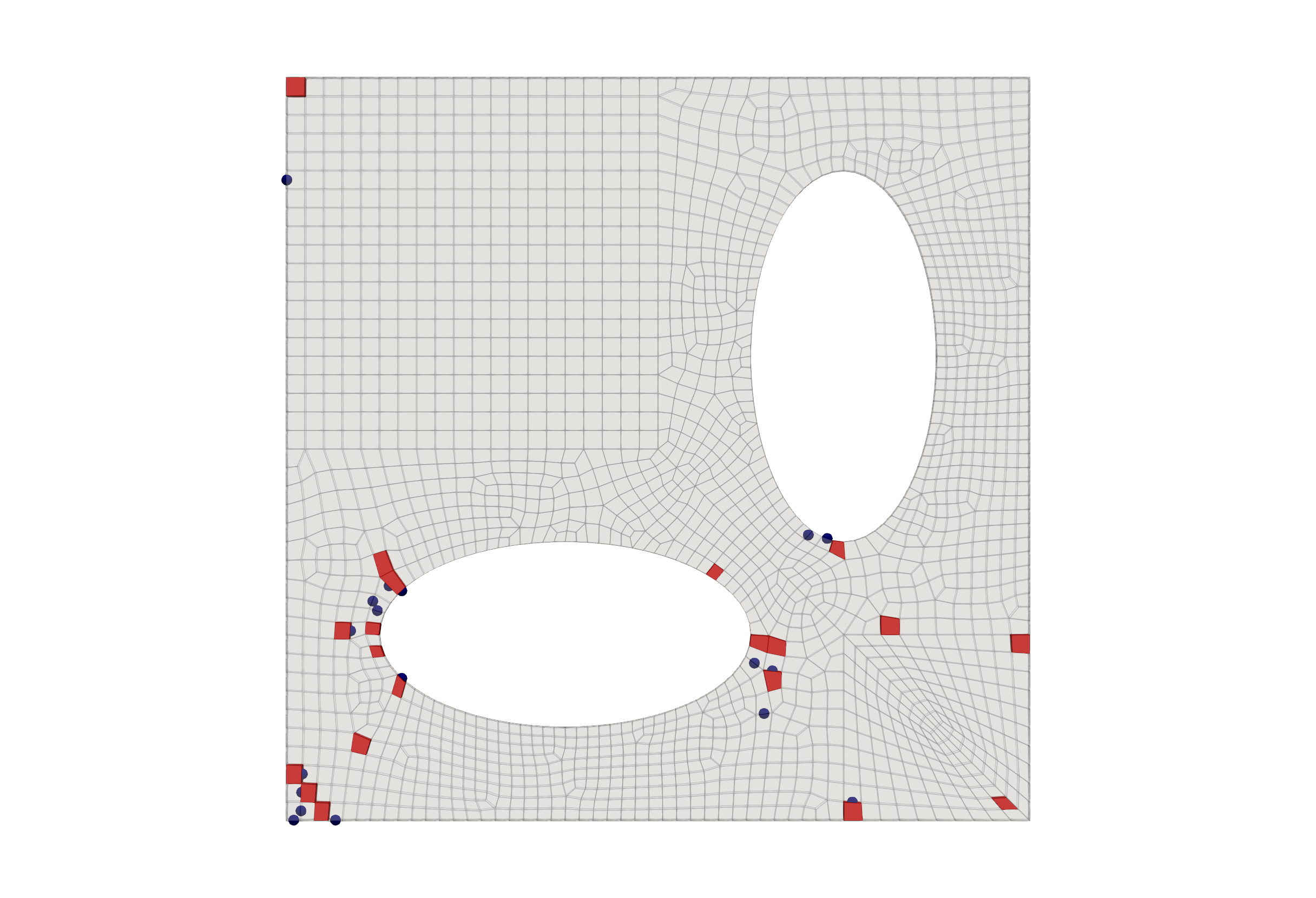}
    \caption{Elements selected by displacement-space ECSW (shaded in red) and Gauss points selected by strain-space ECM (blue dots), for $d=5,m=20$.}
    \label{fig:ECSW_ECM_RID}
\end{figure}

To illustrate how hyperreduced integration is performed on the example problem more specifically, Fig.~\ref{fig:ECSW_ECM_RID} shows the elements selected by displacement space ECSW (highlighted in red) as well as the integration points chosen by ECM (blue dots), for $d=5,m=20$. As expected, these are concentrated in regions of high strain fluctuation, and the correlation between the locations sampled by the two methods is high. In both cases, the (very sparse) sampling is sufficient for an error level below $1\%$.

\begin{figure}[h]
    \centering
    \includegraphics[width=0.6\linewidth,trim={15cm 3.5cm 12cm 5cm},clip]{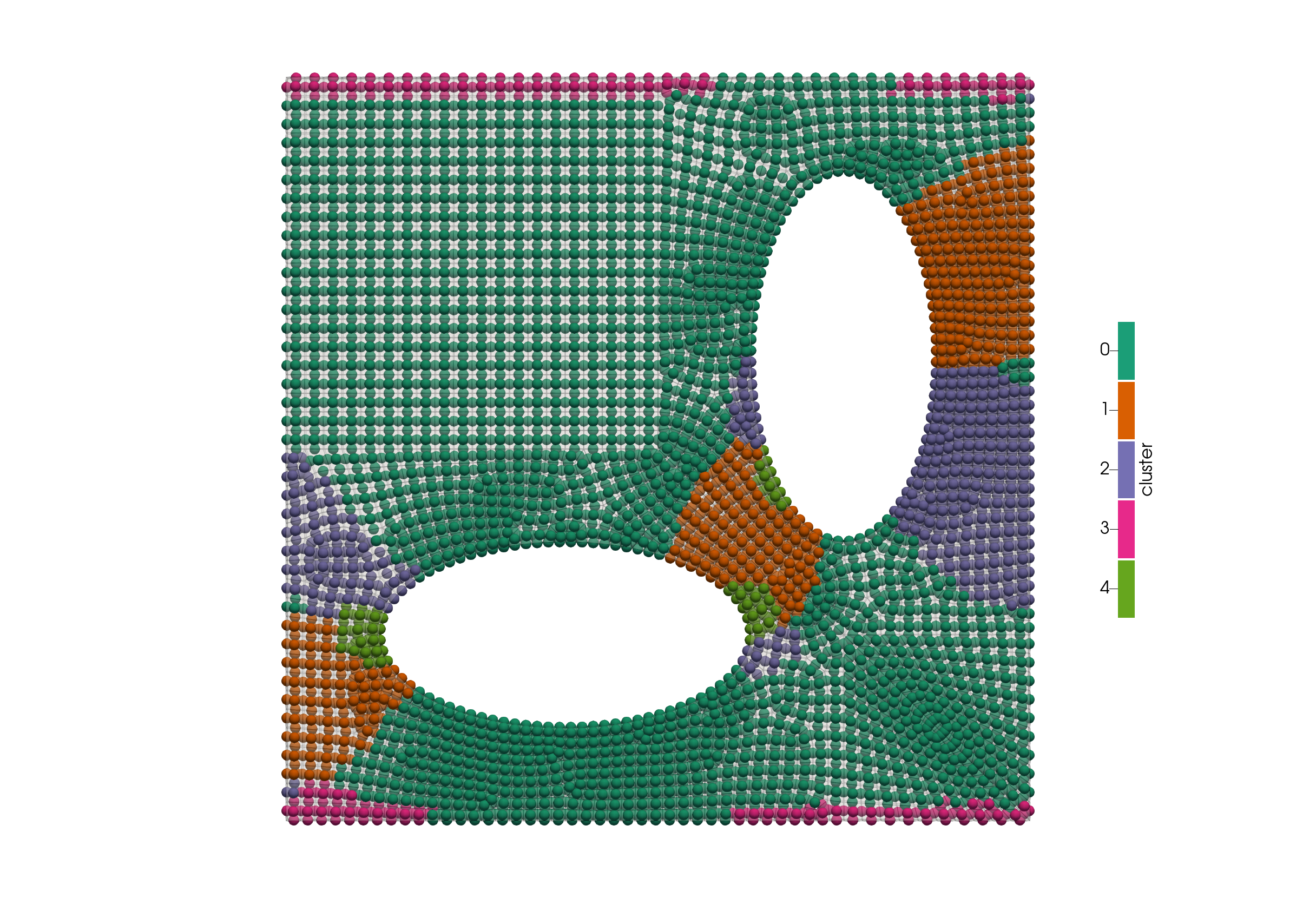}
    \caption{Cluster association of all Gauss point, as chosen by EMSL, for $d=5,m=5$.}
    \label{fig:EMSL_RID}
\end{figure}

Similarly, Fig.~\ref{fig:EMSL_RID} shows the cluster association of each Gauss point when using EMSL with $d=5,m=5$. This parameter combination is sufficient for below $0.25\%$ error. For each cluster, only one evaluation of the material routine (at the expected average cluster strain value) is performed in each load step. While clusters as a whole are not contiguous, contiguous regions of space do tend to be assigned to the same cluster. This is to be expected, because strains often behave similarly in material neighbourhoods. There are some notable exceptions, with only one very few neighbouring Gauss points being associated with the same cluster. This tends to happen in transition regions between different clusters. At several boundaries between clusters 1 and 2, for example, there are transition regions assigned to clusters 0 and 4.
As is to be expected, regions which behave similarly (boundaries with Dirichlet conditions, edges of the ellipse with similar curvature, etc.) are assigned to the same cluster. As a result, the cluster associated with relatively low strains (cluster 0) is comparatively large, and the clusters associated with the largest strains (3 and 4) are much smaller. Because these small regions of large strain have a disproportionate effect on the overall mechanical behaviour, resolving them more finely in the reduced model allows computational resources to be allocated more effectively.

E3C can not easily be visualised on the mesh used in the original FE simulations, because the strain-space optimisation leaves the reduced integration points without counterparts in this mesh. The E3C integration points are however initialised with the strain-space centroids of the Gauss point clusters also utilised by EMSL and shown in Fig.~\ref{fig:EMSL_RID}.

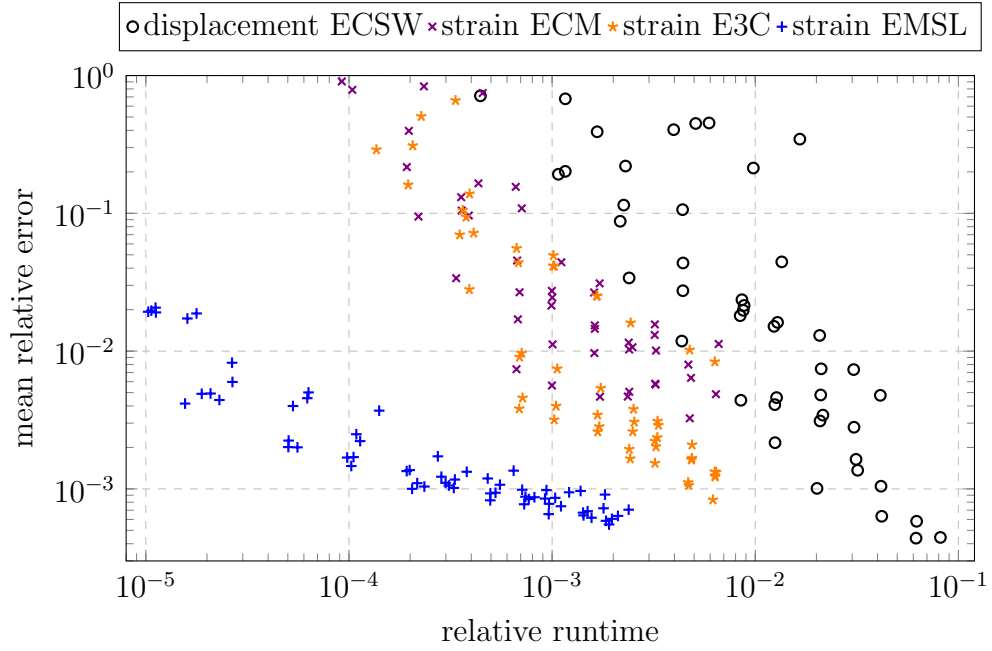
\begin{figure*}[h!]
\centering
    \centering
    \begin{tikzpicture}
    \begin{loglogaxis}[
        ylabel={mean relative error},
        xlabel={relative runtime},
        xmin=0.8e-5, xmax=1.2e-1, domain=0.8e-5:1.2e-1,
        ymin=3e-4, ymax=1e0, domain=3e-4:1e0,
        legend style = {
        at={(0.5,1.05)},
        anchor=south,
        legend columns = -1,
        },
        ymajorgrids=true,
        xmajorgrids=true,
        grid style=dashed,
        table/col sep=semicolon,
        width=0.8\textwidth,
        height=0.5\textwidth,
    ]

    \addplot+[
        black,
        only marks,
        thick,
        mark=o]
    table[x index=4,y index=2] 
    {POD_ECSW_new.csv};
    \addlegendentry{displacement ECSW}

    \addplot+[
        violet,
        only marks,
        thick,
        mark=x]
    table[x index=4,y index=2] 
    {strain_POD_ECM_new.csv};
    \addlegendentry{strain ECM}
    
    \addplot+[
        orange,
        only marks,
        thick,
        mark=star]
    table[x index=4,y index=2] 
    {strain_POD_E3C_new.csv};
    \addlegendentry{strain E3C}

    \addplot+[
        blue,
        only marks,
        thick,
        mark=+]
    table[x index=4,y index=2] 
    {strain_POD_EMSL_new.csv};
    \addlegendentry{strain EMSL}

    \end{loglogaxis}
    \end{tikzpicture}
    \caption{Relative error over relative runtime for displacement-space ECSW (black circles), as well as strain-space ECM (violet $\times$ symbols), E3C (orange stars), and EMSL (blue plus signs), logarithmic scaling for error- and time axes.}
    \label{fig:etlog1}
\end{figure*}

Finally, Fig.~\ref{fig:etlog1} shows the relative mean error level achieved in the validation set over the relative runtime budget required for all validation simulations to complete. Each point in this Figure corresponds to the error/runtime performance achieved by the given hyperreduction technique for one combination of $d$ and $|H|$. Both the relative error on the $y$- and the relative runtime on the $x$-axis are scaled logarithmically. 

When comparing the optimal realisable error-runtime tradeoff, the strain-space variant of ECM and E3C clearly Pareto-dominate displacement-space ECSW, and EMSL Pareto-dominates all competing methods. There is some overlap in the achievable performance range between ECM and E3C, but E3C still has a distinct performance advantage, especially when high levels of accuracy are desired.

In order to realise an error level of $0.1\%$, ECSW requires a runtime budget of around $2\%$ relative to the runtime of the original model, while E3C requires about $0.5\%$ and EMSL only $0.02\%$. At this level of accuracy, E3C thus yields a speedup of almost one order of magnitude over ECSW, while EMSL accelerates simulations by yet another order of magnitude. An error level of $1\%$, meanwhile, can be realised in around $0.4\%$, $0.07\%$, and $0.002\%$ of the original runtime using ECSW, E3C, and EMSL, respectively.

Alternatively, performance can be assessed in terms of the error achievable within a given computational budget. ECSW achieves a $100$-fold runtime reduction with an error of around $0.2\%$. ECM and E3C enable a $1000$-fold speedup with mean relative errors of around $0.6\%$ and $0.3\%$, while EMSL does so with around $0.06\%$ error. A $10,000$-fold speedup is only realistically realisable with EMSL, with below $0.2\%$ relative error.
Whether we consider error and runtime performance at fixed budgets or fixed error levels, in this simple example E3C thus improves on the results obtained by ECSW by roughly one order of magnitude. EMSL improves upon this by about another order of magnitude.

\begin{figure*}[h!]
\begin{subfigure}[t]{.49\textwidth}
  \centering
    \def\svgwidth{\linewidth}
    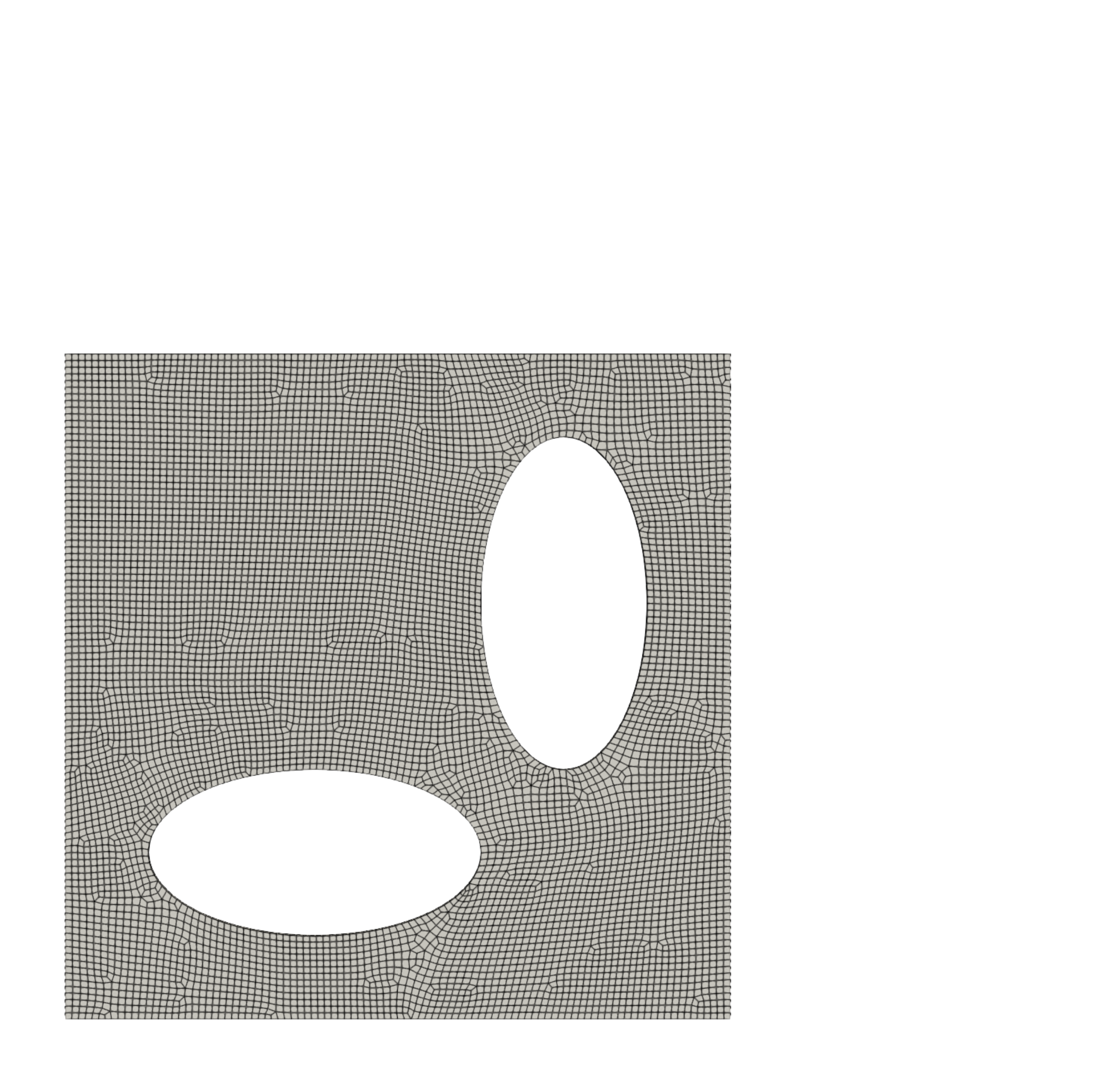
  \caption{Discretisation of plate with two elliptical holes.}
  \label{fig:pw2h}
\end{subfigure}
\hspace{0.05cm}
\begin{subfigure}[t]{.49\textwidth}
  \centering
    \def\svgwidth{\linewidth}
\begingroup%
  \makeatletter%
  \providecommand\color[2][]{%
    \errmessage{(Inkscape) Color is used for the text in Inkscape, but the package 'color.sty' is not loaded}%
    \renewcommand\color[2][]{}%
  }%
  \providecommand\transparent[1]{%
    \errmessage{(Inkscape) Transparency is used (non-zero) for the text in Inkscape, but the package 'transparent.sty' is not loaded}%
    \renewcommand\transparent[1]{}%
  }%
  \providecommand\rotatebox[2]{#2}%
  \newcommand*\fsize{\dimexpr\f@size pt\relax}%
  \newcommand*\lineheight[1]{\fontsize{\fsize}{#1\fsize}\selectfont}%
  \ifx\svgwidth\undefined%
    \setlength{\unitlength}{1124.50927734bp}%
    \ifx\svgscale\undefined%
      \relax%
    \else%
      \setlength{\unitlength}{\unitlength * \real{\svgscale}}%
    \fi%
  \else%
    \setlength{\unitlength}{\svgwidth}%
  \fi%
  \global\let\svgwidth\undefined%
  \global\let\svgscale\undefined%
  \makeatother%
  \begin{picture}(1,0.97043119)%
    \lineheight{1}%
    \setlength\tabcolsep{0pt}%
    \put(0,0){\includegraphics[width=\unitlength,page=1]{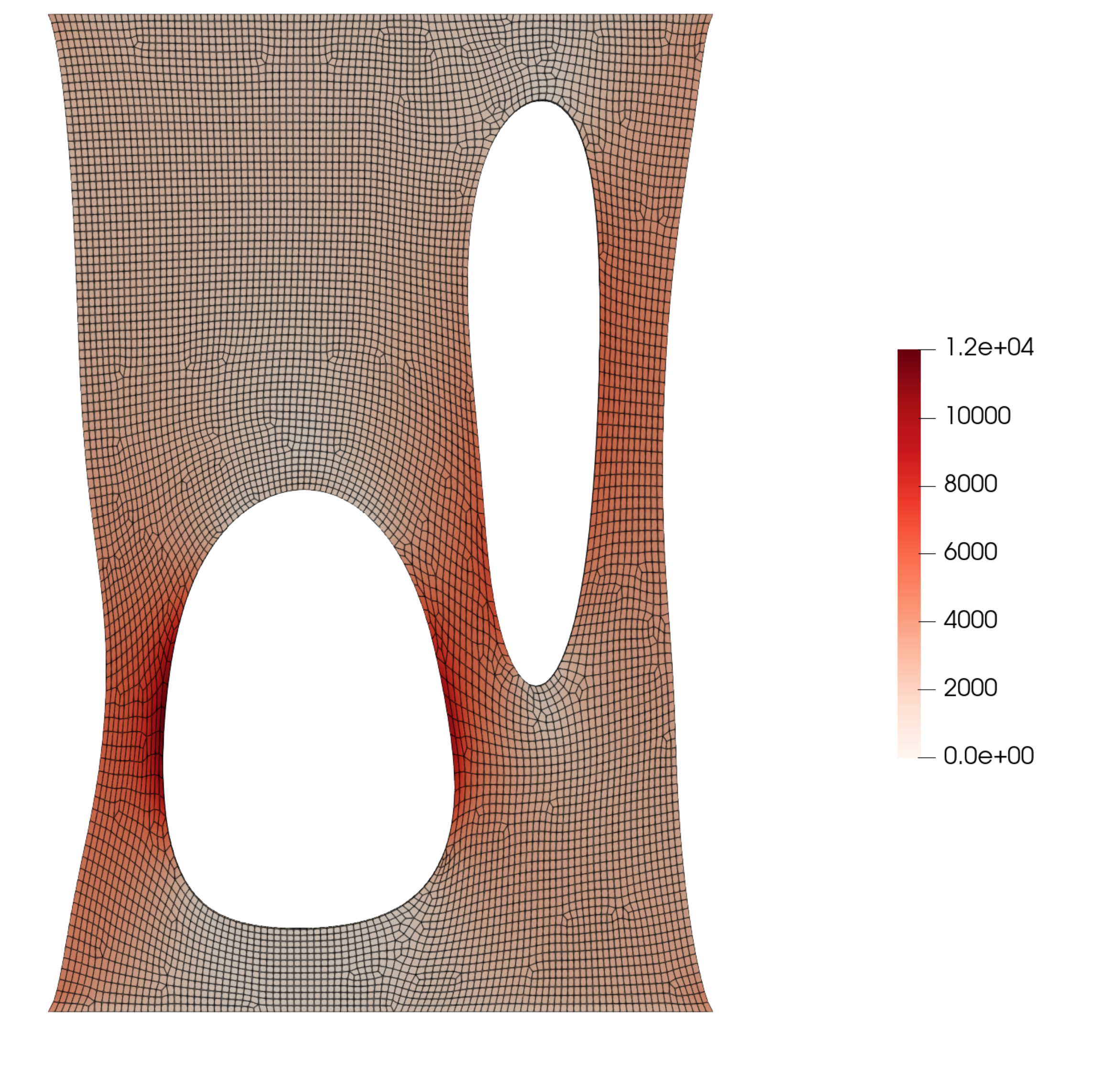}}%
    \put(0.90810278,0.4700562){\color[rgb]{0,0,0}\makebox(0,0)[lt]{\lineheight{1.25}\smash{\begin{tabular}[t]{l}$\sigma^{\text{vM}}$\end{tabular}}}}%
  \end{picture}%
\endgroup%

  \caption{Deformed plate at the end of tensile load path for one parameter sample. Deformation shown to scale, von Mises stress $\sigma^{\text{vM}}$ shown using heatmap.}
  \label{fig:tension}
\end{subfigure}
\vspace{0.2cm}

\begin{subfigure}[t]{.49\textwidth}
  \centering
    \def\svgwidth{\linewidth}
\begingroup%
  \makeatletter%
  \providecommand\color[2][]{%
    \errmessage{(Inkscape) Color is used for the text in Inkscape, but the package 'color.sty' is not loaded}%
    \renewcommand\color[2][]{}%
  }%
  \providecommand\transparent[1]{%
    \errmessage{(Inkscape) Transparency is used (non-zero) for the text in Inkscape, but the package 'transparent.sty' is not loaded}%
    \renewcommand\transparent[1]{}%
  }%
  \providecommand\rotatebox[2]{#2}%
  \newcommand*\fsize{\dimexpr\f@size pt\relax}%
  \newcommand*\lineheight[1]{\fontsize{\fsize}{#1\fsize}\selectfont}%
  \ifx\svgwidth\undefined%
    \setlength{\unitlength}{1124.50927734bp}%
    \ifx\svgscale\undefined%
      \relax%
    \else%
      \setlength{\unitlength}{\unitlength * \real{\svgscale}}%
    \fi%
  \else%
    \setlength{\unitlength}{\svgwidth}%
  \fi%
  \global\let\svgwidth\undefined%
  \global\let\svgscale\undefined%
  \makeatother%
  \begin{picture}(1,0.97043119)%
    \lineheight{1}%
    \setlength\tabcolsep{0pt}%
    \put(0,0){\includegraphics[width=\unitlength,page=1]{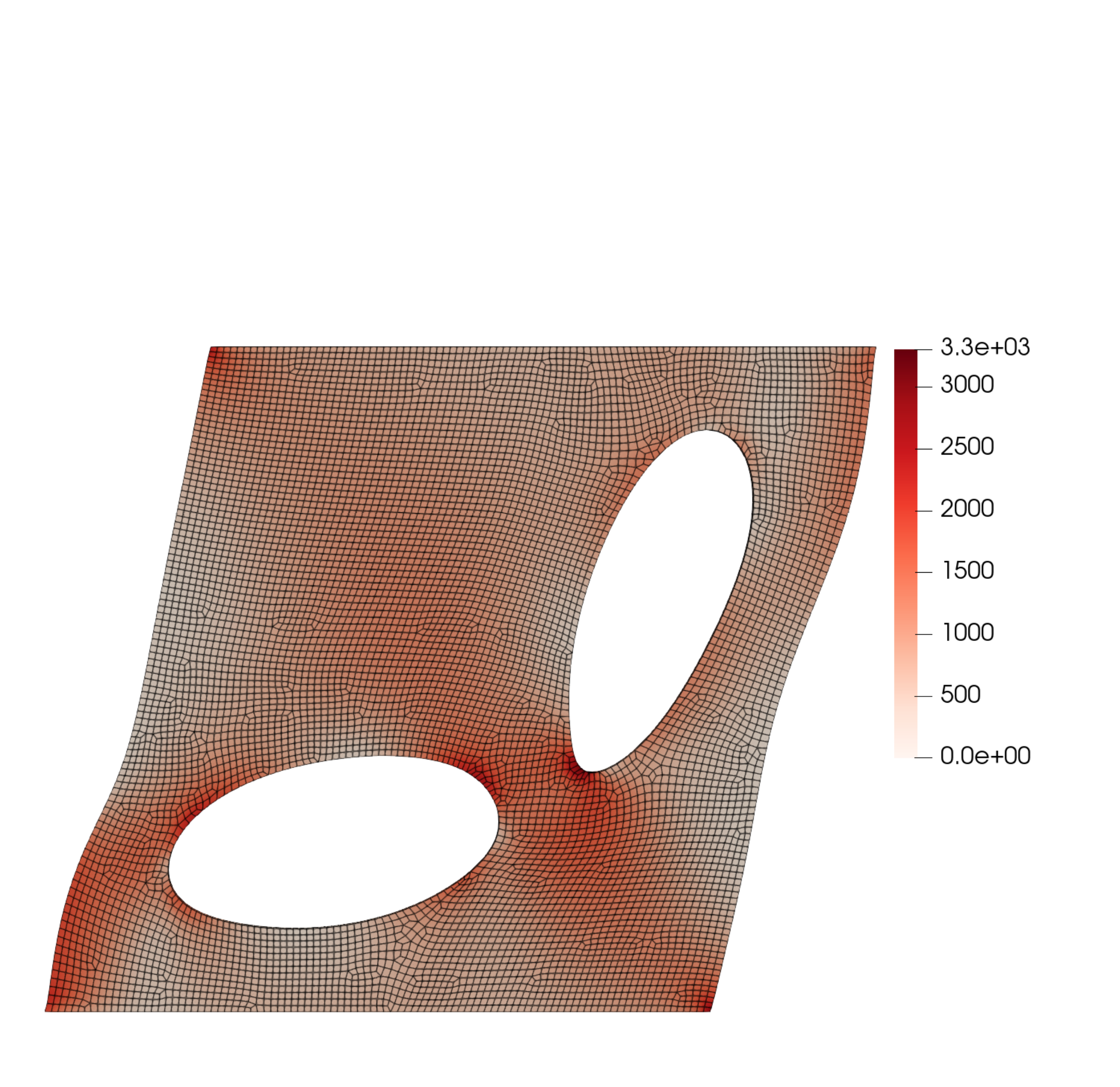}}%
    \put(0.90810286,0.47005622){\color[rgb]{0,0,0}\makebox(0,0)[lt]{\lineheight{1.25}\smash{\begin{tabular}[t]{l}$\sigma^{\text{vM}}$\end{tabular}}}}%
  \end{picture}%
\endgroup%

  \caption{Deformed plate at the end of shear load path for one parameter sample. Deformation shown to scale, von Mises stress $\sigma^{\text{vM}}$ shown using heatmap.}
  \label{fig:shear}
\end{subfigure}
\hspace{0.05cm}
\begin{subfigure}[t]{.49\textwidth}
  \centering
    \def\svgwidth{\linewidth}
\begingroup%
  \makeatletter%
  \providecommand\color[2][]{%
    \errmessage{(Inkscape) Color is used for the text in Inkscape, but the package 'color.sty' is not loaded}%
    \renewcommand\color[2][]{}%
  }%
  \providecommand\transparent[1]{%
    \errmessage{(Inkscape) Transparency is used (non-zero) for the text in Inkscape, but the package 'transparent.sty' is not loaded}%
    \renewcommand\transparent[1]{}%
  }%
  \providecommand\rotatebox[2]{#2}%
  \newcommand*\fsize{\dimexpr\f@size pt\relax}%
  \newcommand*\lineheight[1]{\fontsize{\fsize}{#1\fsize}\selectfont}%
  \ifx\svgwidth\undefined%
    \setlength{\unitlength}{1124.50927734bp}%
    \ifx\svgscale\undefined%
      \relax%
    \else%
      \setlength{\unitlength}{\unitlength * \real{\svgscale}}%
    \fi%
  \else%
    \setlength{\unitlength}{\svgwidth}%
  \fi%
  \global\let\svgwidth\undefined%
  \global\let\svgscale\undefined%
  \makeatother%
  \begin{picture}(1,0.97043119)%
    \lineheight{1}%
    \setlength\tabcolsep{0pt}%
    \put(0,0){\includegraphics[width=\unitlength,page=1]{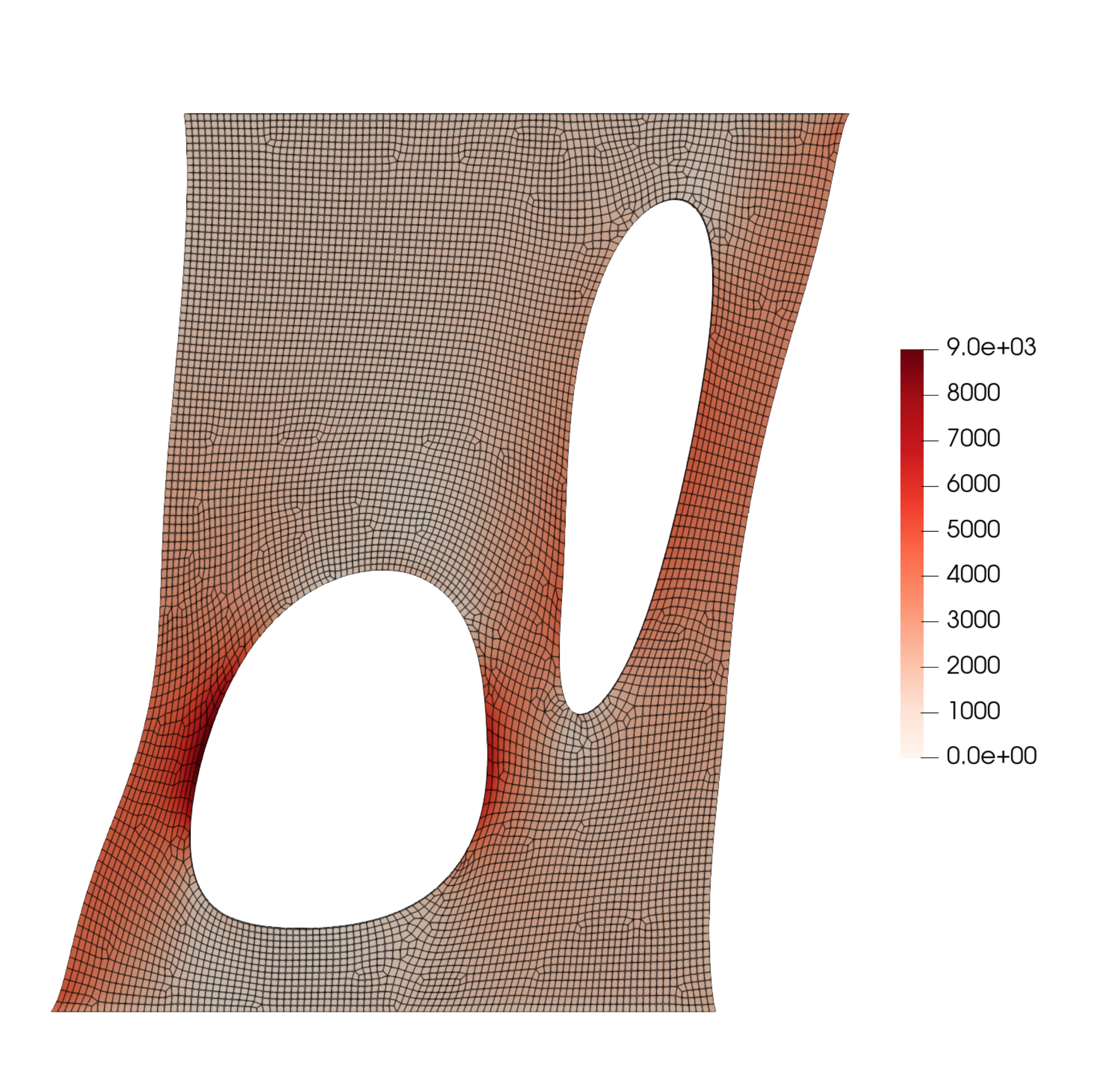}}%
    \put(0.90810282,0.47005622){\color[rgb]{0,0,0}\makebox(0,0)[lt]{\lineheight{1.25}\smash{\begin{tabular}[t]{l}$\sigma^{\text{vM}}$\end{tabular}}}}%
  \end{picture}%
\endgroup%

  \caption{Deformed plate at the end of simultaneous tension and shear load path for one parameter sample. Deformation shown to scale, von Mises stress $\sigma^{\text{vM}}$ shown using heatmap.}
  \label{fig:both}
\end{subfigure}

\caption{Undeformed configuration of plate with two holes and deformation and stress at the end of tensile, shear and simultaneous tension and shear load paths, for one material parameter sample.}
\label{fig:plate}
\end{figure*}
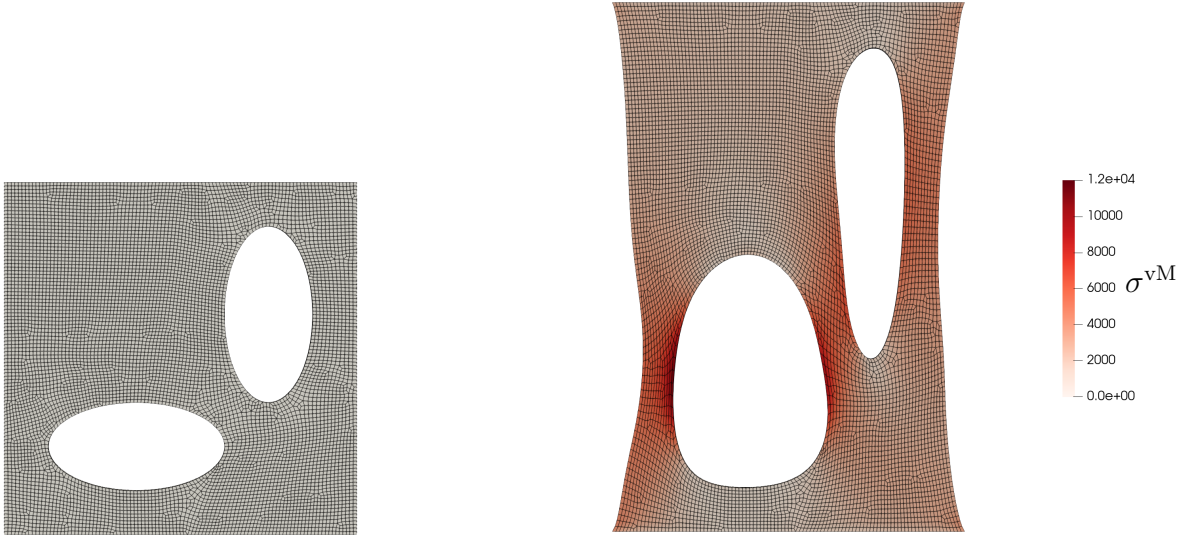
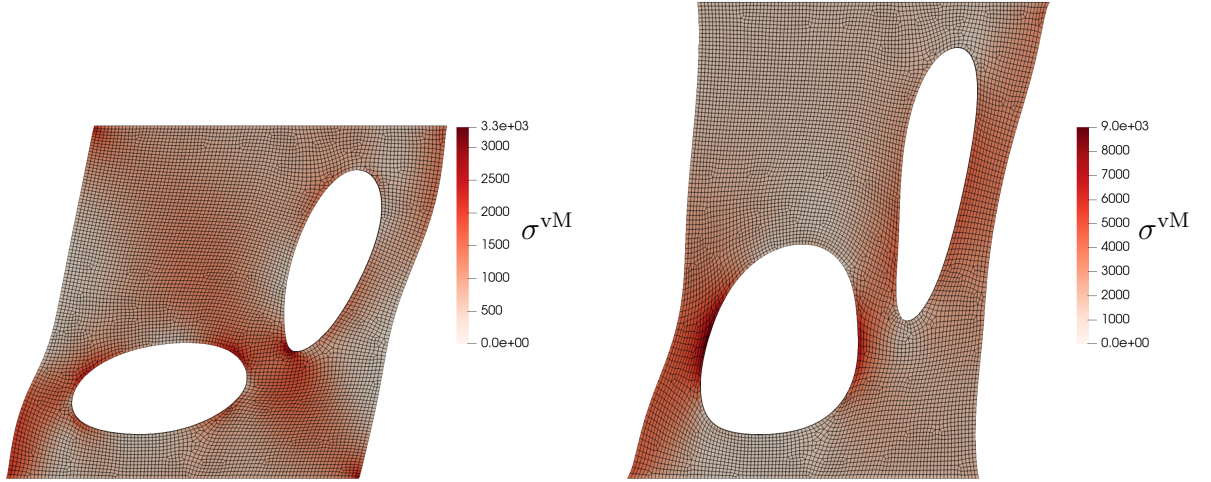

\subsection{Second example: multiple load cases, large deformations, higher computational costs}\label{s:tensionshear}

\subsubsection*{Setup}

In the second example, we employ a finer mesh using $7824$ linear hexahedral elements and $16,288$ nodes, i.e. $48,864$ degrees of freedom. The resulting discretisation is shown in Fig.~\ref{fig:pw2h}. While this second benchmark problem is considerably larger than that explored in Subsection~\ref{s:tension} in a numerical sense, assembly costs are still relevant here, meaning that an efficient hyperreduction is necessary to accelerate simulations significantly.

The plate with two holes is now subjected to three different load cases. Firstly, a tensile Dirichlet BC is applied to the upper surface, with $d_1 \in [0,50]$, i.e. up to $50\%$ average strain. Local strains exceed $150\%$. The displacement resulting from this load case is shown, for an example material parameter combination, in Fig.~\ref{fig:tension}.
Secondly, we apply a shear Dirichlet BC with $d_2 \in [0,25]$ to the same surface, as shown in Fig.~\ref{fig:shear}. Finally, we apply tension and shear simultaneously, defining a proportional load path with $d_1 \in [0,35]$ and $d_2 \in [0,20]$. A displacement result for these boundary conditions is shown in Fig.~\ref{fig:both}.
$10$ equally spaced load steps are scheduled for each load path, but the step size is halved if the Newton-Raphson solver diverges. Three separate loading scenarios are considered to illustrate that the proposed boundary handling strategy is not limited to single Dirichlet boundary conditions.

As above, we employ the compressible Mooney-Rivlin material law in Eq.~\eqref{eq:MR} and vary the material parameters in ranges of $\lambda \in [400,10000]$, and $c_1, c_2 \in [100,1000]$, respectively.
The $25$ training and validation samples for the material parameters are the same as those used for the prior example, as shown in Fig.~\ref{fig:samples}. Since there are $10$ load steps per load path and $3$ load paths per sample now, we have $s=750$ snapshots, and $750$ independent validation parameter combinations in the same parameter range.

For boundary handling, we now compute two solutions to serve as BC-consistent fields, with boundary values of $d_1=50$ and $d_2=25$, respectively. The solutions are scaled by $\frac{1}{50}$ and $\frac{1}{25}$ so that the values at the Dirichlet boundaries become $1$. The reference material parameters $\bar{\lambda}$, $\bar{c}_1$, and $\bar{c}_2$ are again chosen to lie in the centre of the sampled parameter range.  For the final load path, which features both tension and shear loading, both BC-consistent fields are activated simultaneously. 

As above, relative errors attained by all MOR techniques are calculated over all $750$ validation simulations according to Eq.~\eqref{eq:error}, and the runtimes required for all $750$ validation simulations are recorded. 
With the same soft- and hardware setup as in Subsection~\ref{s:tension}, the runtime required for the high-fidelity model to produce solutions for the new validation set is $168,983$~s (approximately $47$ hours) of which $0.07888$ s is overhead incurred due to sampling. 
Additionally, we also measure the time incurred by each method in the offline training phase. 
Of course, the aforementioned caveats concerning runtime measurements still remain valid.

To explore the accuracy/runtime tradeoff achievable with a variety of algorithmic parameters, simulations using all MOR techniques are performed for a range of reduced basis sizes $d\in [5,30]$ and reduced integration points $|H|\in[1,100]$. In ECSW, we vary the selected elements $|E_\text{m}|\in[1,100]$, i.e. $|H|\in[6,600]$.

\subsubsection*{Results}

\begin{figure}[h]
    \centering
    \begin{tikzpicture}
    \begin{semilogyaxis}[
        ylabel={mean relative error},
        xlabel={number of integration points $|H|$},
        xmin=0, xmax=100,
        ymin=3e-4, ymax=1e0, domain=3e-4:1e0,
        xtick={0,5,10,20,30,50,75,100},
        legend pos=outer north east,
        ymajorgrids=true,
        xmajorgrids=true,
        grid style=dashed,
        table/col sep=semicolon,
    ]

    \addplot+[
        black,
        semithick,
        mark=o]
    table[x index=5,y index=1] 
    {allemd10.csv};
    \addlegendentry{displacement POD-ECSW}

    \addplot+[
        violet,
        semithick,
        mark=x]
    table[x index=0,y index=2] 
    {allemd10.csv};
    \addlegendentry{strain POD-ECM}

    \addplot+[
        orange,
        semithick,
        mark=star]
    table[x index=0,y index=3] 
    {allemd10.csv};
    \addlegendentry{strain POD-E3C}
    
    \addplot+[
        blue,
        semithick,
        mark=+]
    table[x index=0,y index=4] 
    {allemd10.csv};
    \addlegendentry{strain POD-EMSL}

    \end{semilogyaxis}
    \end{tikzpicture}
    \caption{Mean relative error over number of integration points $|H|$, for $d=10$ and displacement-space ECSW (black circles), as well as strain-space ECM (violet $\times$ symbols), E3C (orange stars), and EMSL (blue plus signs).}
    \label{fig:ed10}
\end{figure}
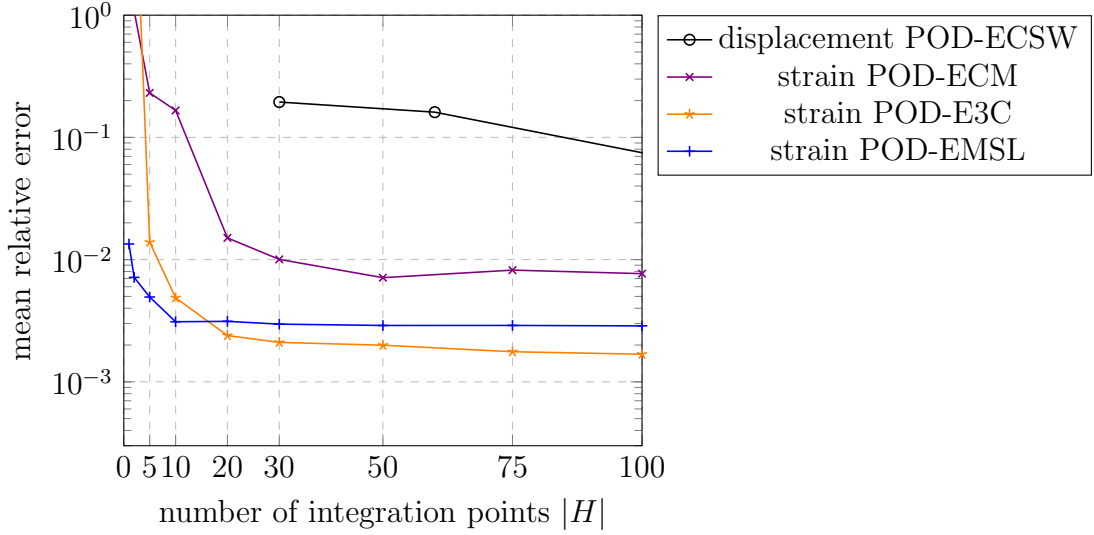

Fig.~\ref{fig:ed10} again shows the mean relative displacement error over a range of integration points $|H|\in [1,100]$, for $d=10$, and all MOR techniques. Once more, the results obtained using ECSW are also plotted over the number of selected integration points $|H|$, not the number of selected elements $|E_\text{m}|$. 

As in the tensile example, the error falls with increasing $|H|$ for all methods. In the case of displacement-space ECSW, it drops gradually throughout the range shown, reaching around $5\%$ at $|H|\approx100$. Eventually, values of around $0.3\%$ are reached at $|H|=600$, $|E_\text{m}|=100$, which lies beyond the range of the plot shown here.
Because ECM allocates integration points more efficiently, it improves on the performance of ECSW over the whole range of $|H|$. The error drops to $1\%$ at $|H|=30$, before stagnating at around $0.7\%$. In this more complicated example and in the parameter range shown, performing the integration point selection on the Gauss point level thus boosts the accuracy achievable with a given number of reduced integration points by one order of magnitude.

E3C is yet more accurate, achieving an error level of below $2\%$ with only $5$ integration points, and below $0.5\%$ with $|H|=10$. 
The error then decreases more slowly to $0.1682\%$ at $|H|=100$. Once more, the optimised integration point selection facilitates significantly higher levels of accuracy, even with very few reduced integration points.

The error achieved by EMSL drops even more rapidly at first, with below $2\%$ being possible using only $1$ reference point and below $0.5\%$ requiring only $|H|=5$. The behaviour becomes asymptotic earlier than in the case of E3C, stagnating at about $0.3\%$ with $|H|=30$ reference points. In this metric, EMSL therefore outperforms ECSW and ECM across the entire investigated parameter range. With $|H|<20$, the performance advantage spans one order of magnitude. It also does better than E3C when few integration points are used, but E3C yields more accurate results for $|H|\geq 20$.

In this more nonlinear example, the inexact integration of the exact material law favoured by E3C (that is, using optimised strain-space integration points) therefore yields more accurate results than an EMSL-like scheme, which exactly integrates an approximation of the material law, provided sufficiently many integration points $|H|\geq 20$ are used. When fewer reduced integration points are used, or when the selection of integration points is constrained to a subset of the original mesh (as in the case of ECSW and ECM), the EMSL-like scheme retains its advantage. The classical Newton-Raphson scheme and Gauss-like strain-space integration employed by E3C seem to benefit more from the additional information contained in additional reduced integration points, while EMSLs advantage lies in extracting relevant information from fewer reference points.

\begin{table}[h!]
    \centering
    \begin{tblr}{
    colspec={cc|ccccccccccc},
cell{3}{3} = {red!30!white},
cell{3}{6} = {red!30!white},
cell{3}{7} = {red!30!white},
cell{3}{8} = {red!30!white},
cell{3}{9} = {purple!30!white},
cell{3}{10} = {violet!30!white},
cell{3}{11} = {violet!30!white},
cell{4}{5} = {red!30!white},
cell{4}{6} = {red!30!white},
cell{4}{7} = {purple!30!white},
cell{4}{8} = {violet!30!white},
cell{4}{9} = {blue!30!white},
cell{4}{10} = {blue!30!white},
cell{4}{11} = {blue!30!white},
cell{5}{5} = {red!30!white},
cell{5}{6} = {red!30!white},
cell{5}{7} = {red!30!white},
cell{5}{8} = {red!30!white},
cell{5}{9} = {blue!30!white},
cell{5}{10} = {blue!30!white},
cell{5}{11} = {blue!30!white},
cell{6}{6} = {red!30!white},
cell{6}{7} = {red!30!white},
cell{6}{8} = {red!30!white},
cell{6}{9} = {violet!30!white},
cell{6}{10} = {blue!30!white},
cell{6}{11} = {blue!30!white},
cell{7}{6} = {red!30!white},
cell{7}{7} = {red!30!white},
cell{7}{8} = {red!30!white},
cell{7}{9} = {violet!30!white},
cell{7}{10} = {violet!30!white},
cell{7}{11} = {cyan!30!white},
cell{8}{5} = {red!30!white},
cell{8}{7} = {red!30!white},
cell{8}{9} = {purple!30!white},
cell{8}{10} = {violet!30!white},
cell{8}{11} = {cyan!30!white},
}   
        & & \SetCell[c=9]{c} $|E_\text{m}|$ & \\
        & & 1 & 2 & 5 & 10 & 20 & 30  & 50 & 75 & 100  \\
        \hline
        \SetCell[r=6]{l} $d$ 
 & 5 & 36.62 & $\times$  & $\times$  & 8.299 & 3.166 & 2.267 & 1.580 & 0.6123 & 0.6195 \\
 & 7 & $\times$  & $\times$  & 8.277 & 3.631 & 1.457 & 0.9989 & 0.4017 & 0.3056 & 0.2379 \\
 & 10 & $\times$  & $\times$  & 19.47 & 16.09 & 5.113 & 2.833 & 0.4280 & 0.3928 & 0.3207 \\
 & 15 & $\times$  & $\times$  & $\times$  & 15.67 & 3.107 & 2.892 & 0.7818 & 0.2507 & 0.2317 \\
 & 20 & $\times$  & $\times$  & $\times$  & 19.35 & 3.904 & 2.388 & 0.5259 & 0.5169 & 0.0984 \\
 & 30 & $\times$  & $\times$  & 13.01 & $\times$  & 12.65 & $\times$  & 1.828 & 0.5278 & 0.1929 \\
    \end{tblr}
    \caption{Relative error ($\%$) obtained using displacement-space ECSW models with different $d$ and $|E_\text{m}|$, relative to original FE simulation.}
    \label{tab:ECSW_error}
\end{table}

\begin{table}[h!]
    \centering
    \begin{tblr}{
    colspec={cc|ccccccccccc},
cell{3}{3} = {red!30!white},
cell{3}{5} = {red!30!white},
cell{3}{6} = {red!30!white},
cell{3}{7} = {red!30!white},
cell{3}{8} = {red!30!white},
cell{3}{9} = {red!30!white},
cell{3}{10} = {red!30!white},
cell{3}{11} = {red!30!white},
cell{4}{4} = {red!30!white},
cell{4}{6} = {red!30!white},
cell{4}{7} = {red!30!white},
cell{4}{8} = {purple!30!white},
cell{4}{9} = {purple!30!white},
cell{4}{10} = {purple!30!white},
cell{4}{11} = {purple!30!white},
cell{5}{4} = {red!30!white},
cell{5}{5} = {red!30!white},
cell{5}{6} = {red!30!white},
cell{5}{7} = {purple!30!white},
cell{5}{8} = {purple!30!white},
cell{5}{9} = {violet!30!white},
cell{5}{10} = {violet!30!white},
cell{5}{11} = {violet!30!white},
cell{6}{6} = {red!30!white},
cell{6}{7} = {red!30!white},
cell{6}{8} = {purple!30!white},
cell{6}{9} = {violet!30!white},
cell{6}{10} = {violet!30!white},
cell{6}{11} = {blue!30!white},
cell{7}{7} = {red!30!white},
cell{7}{8} = {purple!30!white},
cell{7}{9} = {purple!30!white},
cell{7}{10} = {violet!30!white},
cell{7}{11} = {purple!30!white},
cell{8}{7} = {red!30!white},
cell{8}{8} = {red!30!white},
cell{8}{9} = {red!30!white},
cell{8}{10} = {purple!30!white},
cell{8}{11} = {violet!30!white},    }   
        & & \SetCell[c=9]{c} $|H|$ & \\
        & & 1 & 2 & 5 & 10 & 20 & 30  & 50 & 75 & 100 \\
        \hline
        \SetCell[r=6]{l} $d$ 
 & 5 & 12.80 & $\times$ & 23.22 & 13.06 & 2.237 & 2.226 & 2.210 & 2.256 & 2.249 \\
 & 7 & $\times$ & 30.73 & $\times$ & 5.801 & 2.395 & 1.382 & 1.276 & 1.218 & 1.199 \\
 & 10 & $\times$ & 108.56 & 23.12 & 16.62 & 1.505 & 1.004 & 0.7115 & 0.8205 & 0.7676 \\
 & 15 & $\times$ & $\times$ & $\times$ & 16.48 & 6.406 & 1.285 & 0.9728 & 0.6558 & 0.4202 \\
 & 20 & $\times$ & $\times$ & $\times$ & $\times$ & 11.08 & 1.336 & 1.257 & 0.7208 & 1.159 \\
 & 30 & $\times$ & $\times$ & $\times$ & $\times$ & 20.49 & 7.826 & 3.522 & 1.345 & 0.7729 \\
    \end{tblr}
    \caption{Relative error ($\%$) obtained using strain-space ECM models with different $d$ and $|H|$, relative to original FE simulation.}
    \label{tab:ECM_error}
\end{table}

\begin{table}[h!]
    \centering
    \begin{tblr}{
    colspec={cc|ccccccccccc},
cell{3}{3} = {red!30!white},
cell{3}{4} = {red!30!white},
cell{3}{5} = {purple!30!white},
cell{3}{6} = {violet!30!white},
cell{3}{7} = {violet!30!white},
cell{3}{8} = {violet!30!white},
cell{3}{9} = {violet!30!white},
cell{3}{10} = {violet!30!white},
cell{3}{11} = {violet!30!white},
cell{4}{5} = {violet!30!white},
cell{4}{6} = {blue!30!white},
cell{4}{7} = {blue!30!white},
cell{4}{8} = {blue!30!white},
cell{4}{9} = {blue!30!white},
cell{4}{10} = {cyan!30!white},
cell{4}{11} = {cyan!30!white},
cell{5}{4} = {red!30!white},
cell{5}{5} = {purple!30!white},
cell{5}{6} = {blue!30!white},
cell{5}{7} = {blue!30!white},
cell{5}{8} = {blue!30!white},
cell{5}{9} = {cyan!30!white},
cell{5}{10} = {cyan!30!white},
cell{5}{11} = {cyan!30!white},
cell{6}{6} = {violet!30!white},
cell{6}{7} = {blue!30!white},
cell{6}{8} = {blue!30!white},
cell{6}{9} = {cyan!30!white},
cell{6}{10} = {cyan!30!white},
cell{6}{11} = {cyan!30!white},
cell{7}{6} = {purple!30!white},
cell{7}{7} = {blue!30!white},
cell{7}{8} = {blue!30!white},
cell{7}{9} = {blue!30!white},
cell{7}{10} = {cyan!30!white},
cell{7}{11} = {cyan!30!white},
cell{8}{7} = {violet!30!white},
cell{8}{8} = {violet!30!white},
cell{8}{9} = {blue!30!white},
cell{8}{10} = {cyan!30!white},
cell{8}{11} = {cyan!30!white},
}   
        & & \SetCell[c=9]{c} $|H|$ & \\
        & & 1 & 2 & 5 & 10 & 20 & 30  & 50 & 75 & 100 \\
        \hline
        \SetCell[r=6]{l} $d$ 
 & 5 & 117.5 & 9.079 & 1.102 & 0.6196 & 0.6050 & 0.6254 & 0.6369 & 0.5905 & 0.5842 \\
 & 7 & $\times$ & $\times$ & 0.8257 & 0.4253 & 0.2582 & 0.2271 & 0.2090 & 0.1986 & 0.1995 \\
 & 10 & $\times$ & 2147 & 1.387 & 0.4870 & 0.2392 & 0.2102 & 0.1994 & 0.1763 & 0.1682 \\
 & 15 & $\times$ & $\times$ & $\times$ & 0.9747 & 0.3325 & 0.2183 & 0.1567 & 0.1639 & 0.1277 \\
 & 20 & $\times$ & $\times$ & $\times$ & 1.587 & 0.3620 & 0.3169 & 0.2043 & 0.1606 & 0.1867 \\
 & 30 & $\times$ & $\times$ & $\times$ & $\times$ & 0.7708 & 0.5322 & 0.2650 & 0.1483 & 0.1464 \\
    \end{tblr}
    \caption{Relative error ($\%$) obtained using strain-space E3C models with different $d$ and $|H|$, relative to original FE simulation.}
    \label{tab:E3C_error}
\end{table}

\begin{table}[h!]
    \centering
    \begin{tblr}{
    colspec={cc|ccccccccccc},
cell{3}{3} = {purple!30!white},
cell{3}{4} = {purple!30!white},
cell{3}{5} = {violet!30!white},
cell{3}{6} = {violet!30!white},
cell{3}{7} = {violet!30!white},
cell{3}{8} = {violet!30!white},
cell{3}{9} = {violet!30!white},
cell{3}{10} = {violet!30!white},
cell{3}{11} = {violet!30!white},
cell{4}{3} = {purple!30!white},
cell{4}{4} = {violet!30!white},
cell{4}{5} = {blue!30!white},
cell{4}{6} = {blue!30!white},
cell{4}{7} = {blue!30!white},
cell{4}{8} = {blue!30!white},
cell{4}{9} = {blue!30!white},
cell{4}{10} = {blue!30!white},
cell{4}{11} = {blue!30!white},
cell{5}{3} = {purple!30!white},
cell{5}{4} = {violet!30!white},
cell{5}{5} = {blue!30!white},
cell{5}{6} = {blue!30!white},
cell{5}{7} = {blue!30!white},
cell{5}{8} = {blue!30!white},
cell{5}{9} = {blue!30!white},
cell{5}{10} = {blue!30!white},
cell{5}{11} = {blue!30!white},
cell{6}{3} = {purple!30!white},
cell{6}{4} = {violet!30!white},
cell{6}{5} = {violet!30!white},
cell{6}{6} = {blue!30!white},
cell{6}{7} = {blue!30!white},
cell{6}{8} = {blue!30!white},
cell{6}{9} = {blue!30!white},
cell{6}{10} = {blue!30!white},
cell{6}{11} = {blue!30!white},
cell{7}{3} = {purple!30!white},
cell{7}{4} = {violet!30!white},
cell{7}{5} = {violet!30!white},
cell{7}{6} = {blue!30!white},
cell{7}{7} = {blue!30!white},
cell{7}{8} = {blue!30!white},
cell{7}{9} = {blue!30!white},
cell{7}{10} = {blue!30!white},
cell{7}{11} = {blue!30!white},
cell{8}{3} = {purple!30!white},
cell{8}{4} = {violet!30!white},
cell{8}{5} = {violet!30!white},
cell{8}{6} = {blue!30!white},
cell{8}{7} = {blue!30!white},
cell{8}{8} = {blue!30!white},
cell{8}{9} = {blue!30!white},
cell{8}{10} = {blue!30!white},
cell{8}{11} = {red!30!white},    }   
        & & \SetCell[c=9]{c} $|H|$ & \\
        & & 1 & 2 & 5 & 10 & 20 & 30  & 50 & 75 & 100 \\
        \hline
        \SetCell[r=6]{l} $d$ 
 & 5 & 1.762 & 1.009 & 0.7295 & 0.7234 & 0.7179 & 0.7031 & 0.7037 & 0.7002 & 0.6973 \\
 & 7 & 1.421 & 0.6766 & 0.4218 & 0.3456 & 0.3414 & 0.3437 & 0.3457 & 0.3456 & 0.3447 \\
 & 10 & 1.343 & 0.7157 & 0.4942 & 0.3102 & 0.3125 & 0.2965 & 0.2892 & 0.2894 & 0.2870 \\
 & 15 & 1.415 & 0.8389 & 0.5163 & 0.3560 & 0.2977 & 0.2703 & 0.2728 & 0.2614 & 0.2587 \\
 & 20 & 1.442 & 0.8665 & 0.6253 & 0.3415 & 0.3368 & 0.2950 & 0.2861 & 0.2716 & 0.3537 \\
 & 30 & 1.461 & 0.8958 & 0.6449 & 0.3720 & 0.3661 & 0.3029 & 0.3121 & 0.2990 & 2.386 \\
    \end{tblr}
    \caption{Relative error ($\%$) obtained using strain-space EMSL models with different $d$ and $|H|$, relative to original FE simulation.}
    \label{tab:EMSL_error}
\end{table}

The trends which can be observed in Tabs.~\ref{tab:ECSW_error},~\ref{tab:ECM_error},~\ref{tab:E3C_error}, and~\ref{tab:EMSL_error} are generally similar to those observed in Fig.~\ref{fig:ed10} for $d=10$. Errors generally drop with $|H|$, albeit not perfectly uniformly. The reduction in the error levels with increasing $d$ is less pronounced than that with $|H|$. 
As seen in Tabs.~\ref{tab:ECSW_error},~\ref{tab:ECM_error},~\ref{tab:E3C_error}, and~\ref{tab:EMSL_error}, there is a drop from $d=5$ to $d=15$ or $d=20$, but further increases in $d$ do not make a significant difference. 
This suggests that the snapshots lie close to a low ($\approx15$)-dimensional linear subspace of the solution space, such that only few POD modes are necessary to accurately represent the solution manifold. The $\delta=5$-dimensional solution manifold seems to feature non-negligible curvature in only a few ($\approx 10$) dimensions. Putting it less precisely, the problem is only moderately, not strongly, nonlinear\footnote{
One way to pragmatically think about the degree of nonlinearity of a solution manifold, closely related to the Kolmogorov n-width~\cite{Peh:2022:bkb,BarFarMad:2023:npma}, is as follows: 
we can imagine a $\delta=5$-dimensional linear reference manifold, for example the tangent space to the solution manifold at some reference point. Here, this reference point might be the reference material parameters $\bar{\lambda}, \bar{c}_1, \bar{c}_2$ and zero load. The tangent space is spanned by the perturbations of the solution due to perturbations of all parameters. In order to map this reference manifold to the solution manifold (with some acceptable degree of error), we would need to introduce curvature in some $d^*-\delta$ (mutually orthogonal) normal directions. The gap $d^*-\delta$ measures by how much the dimensionality of a linear approximation space needs to exceed the dimensionality of the solution manifold, if the system behaviour is to be captured up to this acceptable degree of error.}.

While the error levels attained by ECSW with a given number of selected mesh entities (elements $E_\text{m}$ in the case of ECSW) can compete with those of ECM, this is no longer the case when viewed relative to the number of integration points $H$ which need to be queried, as discussed in the context of Fig.~\ref{fig:ed10}. Interestingly, for a given number of selected mesh entities ($|E_\text{m}|$ for ECSW vs. $|H|$ for ECM), ECSW is often slightly, but not significantly, more accurate. This implies that the additional Gauss points that are queried in each selected ECSW element do not contribute much additional information over that contained in a single Gauss point chosen by ECM. This is unsurprising, because the strains in different Gauss points in the same element correlate strongly. 
Again, the more efficient selection of integration points $H$ rather than elements for reduced integration therefore facilitates a more efficient allocation of computational resources and higher levels of accuracy. 

When comparing the strain-space methods, E3C and EMSL can be observed to outperform ECM by a notable margin. EMSL is able to produce accurate results with very few reference points, while E3C achieves even higher levels of accuracy when comparatively many integration points are used. For example, at $d=7$, ECM never achieves an error level of below $0.5\%$ (ECSW needs $|H|=300$ Gauss points to do so), while E3C and EMSL require only $10$ and $5$ integration points to this end, respectively. An error of below $0.2\%$ can only be achieved by E3C, with $|H|=75$ (and ECSW, but with $|E_\text{m}|=100, |H|=600$). ECM generally struggles to yield highly accurate results in the investigated parameter range, only achieving an error below $0.5\%$ on one occasion. All other methods are able to produce results with below $0.3\%$ error on several occasions, though, as mentioned above, we show results for ECSW up to $|E_\text{m}|=100$, i.e. $|H|=600$, which flatters this method in that regard. 

All hyperreduction techniques perform quite robustly. In all parameter combinations with $|H|>d$, all 750 validation simulations converge for all methods, and trends in the error are generally predictable. Increasing $|H|$ increases accuracy up to a point, when stagnation occurs, and increasing $d$ generally increases accuracy up to $d=15$. There is one outlier in the case of EMSL, with $d=30, |H|=100$. This is due to a dramatically wrong simulation with $1580\%$ error because of an overfitting-like effect as clusters become very small. Because this parameter combination lies far beyond the optimal parameter range (EMSL is best used with very few clusters) and because the failure can be caught by basic sanity checks, the outlier is not too problematic. 

On the other hand, EMSL robustly yields results when very few reference points are used. Other methods appear to require $|H|>d$ integration points to be able to robustly identify search directions in the $d$-dimensional linear approximation space. EMSL is not constrained this way, because the linearisation strategy used for integration results allows the material behaviour at all $|G|$ Gauss points to be estimated. It appears that having an approximation of the behaviour at $|G|$ Gauss points rather than only $|H|\ll |G|$ more precise, but limited samples pays off especially when only few queries to the material law are desired. If the linearisation is sufficiently accurate, the resulting linear problem is trivial to solve robustly.

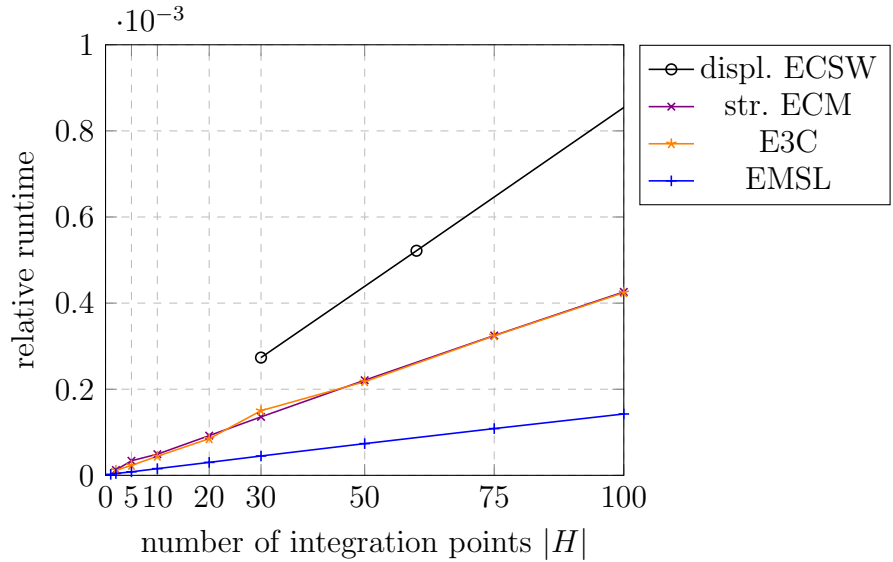
\begin{figure}[h]
    \centering
    \begin{tikzpicture}
    \begin{axis}[
        ylabel={relative runtime},
        xlabel={number of integration points $|H|$},
        xmin=0, xmax=100,
        ymin=0, ymax=0.001,
        xtick={0,5,10,20,30,50,75,100},
        ytick={0,0.0002,0.0004,0.0006,0.0008,0.001},
        legend pos=outer north east,
        ymajorgrids=true,
        xmajorgrids=true,
        grid style=dashed,
        table/col sep=semicolon,
    ]

    \addplot+[
        black,
        semithick,
        mark=o]
    table[x index=5,y index=1] 
    {alltmd10.csv};
    \addlegendentry{displ. ECSW}

    \addplot+[
        violet,
        semithick,
        mark=x]
    table[x index=0,y index=2] 
    {alltmd10.csv};
    \addlegendentry{str. ECM}

    \addplot+[
        orange,
        semithick,
        mark=star]
    table[x index=0,y index=3] 
    {alltmd10.csv};
    \addlegendentry{E3C}
    
    \addplot+[
        blue,
        semithick,
        mark=+]
    table[x index=0,y index=4] 
    {alltmd10.csv};
    \addlegendentry{EMSL}

    \end{axis}
    \end{tikzpicture}
    \caption{Relative runtime over number of integration points $m$, for $d=10$ for displacement-space (black circles), as well as strain-space ECM (violet $\times$ symbols), E3C (orange stars), and EMSL (blue plus signs).}
    \label{fig:dt10}
\end{figure}

Next, Fig.~\ref{fig:dt10} shows the evolution of the relative runtime over the number of integration points $|H|$, again for $d=10$. As in the simpler example, all methods scale close to linearly with $|H|$, as the increase in the number of integration points necessitates proportionately more queries to the material law. 
Once more, the runtimes incurred by ECM and E3C are nearly identical because identical solver algorithms are used in the online phase. Only very slight deviations occur when slightly more or fewer Newton iterations are required. 
Compared to ECM and E3C, ECSW again requires about twice as much runtime due to the additional cost of element-level operations. As in the simpler example above, the reduced complexity of strain-space MOR techniques pays off in the online phase.

Finally, EMSL slashes the runtimes required by ECM and E3C by slightly more than half (as was the case in the simpler example), because iterations are not required within a load step and fewer queries to the material routine are required.
More detailed timing results over the entire range of $d\in[5,30]$ and $|H|\in[1,100]$ can be found in Appendix~\ref{a:timing}, but since the trends visible therein are very similar to those shown in Fig.~\ref{fig:dt10}, we do not discuss these here.

\begin{figure*}[h!]
\centering
    \centering
    \begin{tikzpicture}
    \begin{loglogaxis}[
        ylabel={mean relative error},
        xlabel={relative runtime},
        xmin=1.7e-6, xmax=6e-4, domain=1.7e-6:6e-4,
        ymin=1e-3, ymax=1e-1, domain=1e-3:1e-1,
        legend style = {
        at={(0.5,1.05)},
        anchor=south,
        legend columns = -1,
        },
        ymajorgrids=true,
        xmajorgrids=true,
        grid style=dashed,
        table/col sep=semicolon,
        width=0.8\textwidth,
        height=0.5\textwidth,
    ]

    \addplot+[
        black,
        only marks,
        thick,
        mark=o]
    table[x index=4,y index=2] 
    {POD_ECSW.csv};
    \addlegendentry{displacement ECSW}

    \addplot+[
        violet,
        only marks,
        thick,
        mark=x]
    table[x index=4,y index=2] 
    {strain_POD_ECM.csv};
    \addlegendentry{strain ECM}
    
    \addplot+[
        orange,
        only marks,
        thick,
        mark=star]
    table[x index=4,y index=2] 
    {strain_POD_E3C.csv};
    \addlegendentry{strain E3C}

    \addplot+[
        blue,
        only marks,
        thick,
        mark=+]
    table[x index=4,y index=2] 
    {strain_POD_EMSL.csv};
    \addlegendentry{strain EMSL}

    \end{loglogaxis}
    \end{tikzpicture}
    \caption{Relative error over relative runtime for displacement-space ECSW (black circles), as well as strain-space ECM (violet $\times$ symbols), E3C (orange stars), and EMSL (blue plus signs), logarithmic scaling for error- and time axes.}
    \label{fig:etlog}
\end{figure*}
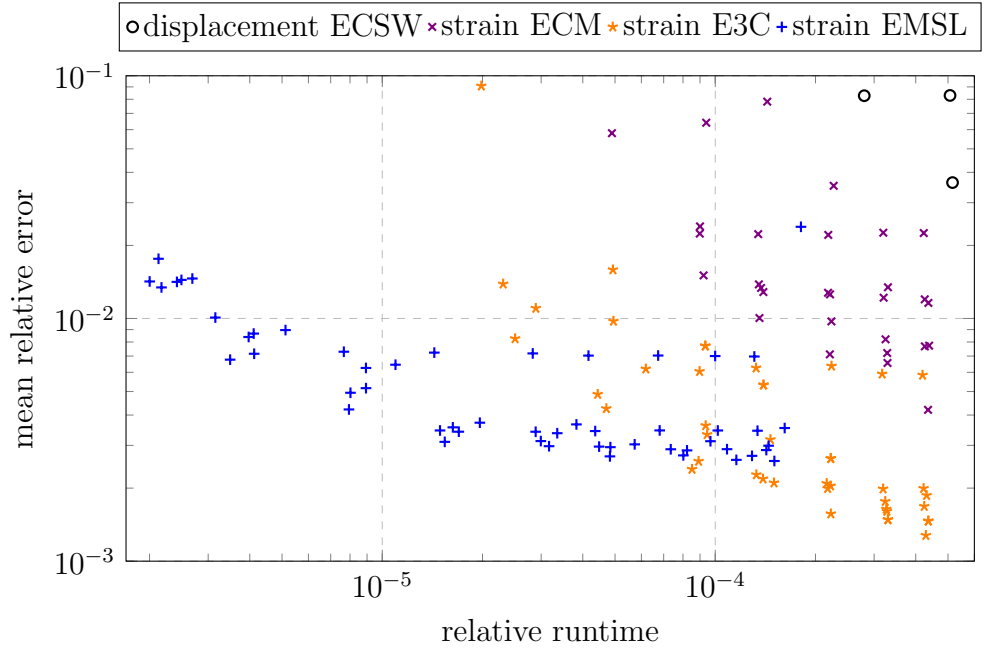

Because the tradeoff between accuracy and runtime is critical to the successful application of MOR schemes, we again show Pareto-curves for all methods in terms of these metrics in Fig.~\ref{fig:etlog}.
The Figure shows the range of error levels and runtimes achievable over all investigated combinations of $d\in[5,30]$ and $|H|\in[1,100]$; the relative error on the $y$- and the relative runtime on the $x$-axis are scaled logarithmically.

When comparing the optimal realisable error-runtime tradeoff, the strain-space methods clearly Pareto-dominate the popular displacement-space technique. Furthermore, E3C and EMSL outperform strain-space ECM by a significant margin. EMSL shines in yielding accurate results in very little runtime, while E3C can produce even more accurate results when the runtime budget is larger. 

In order to realise an error level of $1\%$, for example, ECSW requires about $0.1\%$ of the runtime of the original model, ECM $0.01\%$, E3C $0.003\%$ and EMSL only $0.0003\%$. ECSW thus achieves a $1000$-fold speedup while only incurring $1\%$ error. This performance is already impressive, but at the same level of accuracy, strain-space ECM facilitates a $10,000$-fold acceleration. E3C accelerates simulations around $30,000$-fold and EMSL around $300,000$-fold, all while errors remain around $1\%$. As discussed previously, conclusions about runtimes need to be made carefully, but the magnitude of these computational cost reductions (with accuracy remaining high) do appear to us to suggest that further research on strain-space MOR techniques might be rewarding.

While EMSL yields exceptional runtime accelerations with around $1\%$ error, E3C facilitates even higher levels of accuracy given slightly higher runtime budgets. An error of around $0.13\%$ is possible in $0.04\%$ of the runtime of the original model. This amounts to a $2,500$-fold speedup.

Alternatively, we can measure performance in terms of the error achievable in a given computational budget. A $10,000$-fold speedup is possible with slightly below $2\%$ error using strain-space ECM while EMSL yields around $0.3\%$ and E3C around $0.25\%$. $100,000$-fold speedup is only possible using EMSL, with an error of around $0.45\%$.
In this more challenging example, EMSL remains competitive especially when low computational cost is a priority, while E3C is to be preferred if high levels of accuracy are paramount. 

Finally, offline costs should not be neglected as a factor driving the applicability of MOR schemes. We show the runtime required for offline snapshot data generation and the training phase of each MOR technique in Fig.~\ref{fig:offline}. Generating all training data requires about $47$ hours. We use quite many ($s=750$) snapshots here, which is why computing these takes a long time. In the simpler example explored above, snapshot generation only required $1$ hour and $5$ minutes.
Meanwhile, postprocessing all integration point stresses and strains takes around $3$ hours and the computation of both BC-consistent fields requires about $37$ minutes. 

E3C features the most expensive offline runtime with about $3$ and a half hours on average because of the iterative empirical correction step. For a problem similar to the one discussed here, this is not a significant issue, as the fitting only needs to be performed a few times to select appropriate hyperparameters for some application. Fitting also takes significantly less time than the training data generation, meaning that E3C does not inflate the offline cost by an extreme factor. 
In strongly nonlinear problems which require large POD bases, increase the size of the fitting problem correspondingly, and of course incur larger runtimes for training data generation, this cost might become more problematic. Similarly, a need for additional integration points increases the cost proportionally. However, because the training cost mostly scales with the dimension of the reduced model, not the original FE discretisation, offline cost is likely to stay manageable even for large problems. Only when parameter studies are to be conducted (such as in this work), do these costs become significant.

ECSW incurs an offline training cost of about $3$ hours, mostly due to the need to compute element residuals. This is similarly expensive as computing Gauss points stresses and strains, but less straightforward to extract from black-box solvers. Furthermore, it needs to be repeated at least once per mode number $d$. The setting up and solving of the NNLS problem only takes a few minutes at worst, but this becomes more expensive as the problem size grows. Strain-space ECM is similarly prudent to ECSW (without the need for residual computations), requiring only $4$ and a half minutes for the NNLS problem. Similarly, EMSL training is completed after about $30$ seconds, as its offline phase only features an SVD, a clustering step, and the computation of some products of the POD bases.

\edef\labellist{"$30$ s","$12119$ s","$267$ s","$10468$ s","$168208$ s"}

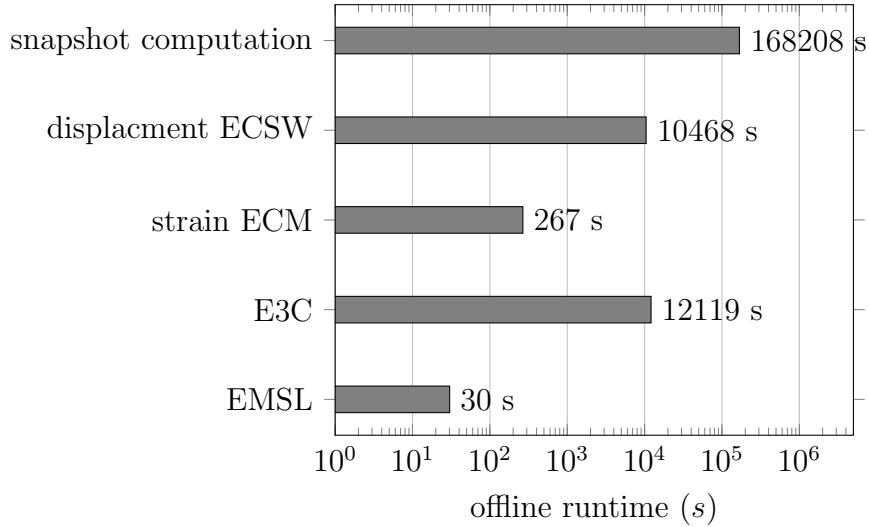
\begin{figure}
    \centering

    \begin{tikzpicture}
        \begin{semilogxaxis}[
            xbar,
            symbolic y coords = {EMSL, E3C, strain ECM, displacment ECSW, snapshot computation},
            ytick={EMSL, E3C, strain ECM, displacment ECSW, snapshot computation},
            nodes near coords=\pgfmathsetmacro{\labelstring}{{\labellist}[\coordindex]}\labelstring,
            xlabel={offline runtime ($s$)},
            xmin=1e0, xmax=5e6, domain=1e0:1e4,
            xmajorgrids=true,
            ]
        \addplot[black,fill=gray] coordinates { (30.22,EMSL) (12119,E3C) (267,strain ECM) (10468,displacment ECSW) (168208,snapshot computation)};
        \end{semilogxaxis}
    \end{tikzpicture}
    
    \caption{Runtimes required for offline snapshot computation and offline training phases of all MOR techniques. Labels rounded to the nearest second.}
    \label{fig:offline}
\end{figure}

\section{Summary and Outlook}

This article generalises strain-space techniques which have recently been shown to achieve exceptional performance in computational homogenisation problems to more general solid-mechanical problems with non-unit-cell geometries. 
To handle parameterised Dirichlet BCs with arbitrary boundary values, we split the displacement gradient field into a BC-consistent reference field and a fluctuation field, similarly to the split often used to handle periodic BCs in the context of RVE problems. This allows us to apply efficient strain-space MOR techniques for computational homogenisation such as E3C to macro-scale, non-RVE problems. Hyperreduction techniques like ECM, which were originally formulated in displacement space and are already popular for structural applications beyond the microscale context, can also be posed in strain space using this boundary handling methodology.

On two hyperelastic example problems motivated by inverse parameter estimation, we observed that strain-space techniques indeed improve on the runtime performance of displacement-space ECSW by a significant margin, as element-level operations are bypassed and queries to the material routine allocated more effectively. Whether viewed from a fixed-budget or fixed-target perspective, E3C outperformed ECSW by more than order of magnitude in these examples. In the second, more challenging example, ECSW facilitated a respectable $1000$-fold speedup at $1\%$ error, but E3C enabled a $30,000$-fold acceleration at the same level of accuracy. Using the same method, a $0.13\%$ error could be achieved in $0.04\%$ of the runtime of the original FE model. 

Additionally, we similarly generalised the EMSL MOR technique, which we recently proposed in the context of computational homogenisation problems. EMSL differs from E3C in that it does not approximately integrate an exact material law, but exactly integrates a cluster-wise linear approximation of the material behaviour around an empirically estimated reference strain. This results in an affine problem in each load increment, doing away with the need for an iterative solver.
This approach did exceptionally well in the first, moderately nonlinear example. For the moderately-sized problem investigated there, EMSL enabled a $1000$-fold runtime acceleration with only around $0.06\%$ error, and a $10,000$-fold acceleration with below $0.2\%$ error. It therefore outperformed ECSW by two orders of magnitude, and E3C by one order of magnitude, whether seen from a fixed-budget or fixed-target perspective, on this example.
Even in the second, more strongly nonlinear problem, a $300,000$-fold speedup was possible with below $1\%$ error. E3C yielded higher levels of accuracy at slightly higher computational cost, but EMSL remained competitive, especially if runtime is prioritised.

The exceptional performance in the example investigated in this work appear to us to justify further research into strain-space MOR techniques beyond the context of computational homogenisation. In addition to the advantages in terms of accuracy and runtime, these strain-space techniques are also attractive on account of their implementational and computational simplicity. Furthermore, only training data -- such as displacements, strains, and stresses -- which can readily be extracted from black-box solvers are required for training. The elimination of element-level operations means that, in addition to this training data, only an appropriate material routine is necessary when deploying the strain-space methods discussed here on a previously unseen parameterised family of problems. For true non-intrusiveness, methods to estimate the material behaviour from stress-strain pairs could also be deployed.

Further research should test the methods outlined here on larger problems. Since the online complexity of these MOR techniques no longer scales with the size of the original discretisation (i.e., $D$ and $|G|$), scaling to very high-dimensional problems should be possible, but more work is required to confirm this. Applications in this range are of course very attractive from a point of view of runtime and memory costs. 

Additionally, extensions to elasto-plastic material behaviour are intriguing, in part because basis splits into elastic and plastic components might be feasible. The strong localisation effects resulting from elasto-plastic behaviour as well as the handling of history behaviour constitute additional challenges for MOR schemes, whether posed in displacement or strain space. In the context of computational homogenisation, Wulfinghoff and Goldbeck have already applied E3C to (small-strain) elasto-plastic behaviour in~\cite{Wul:2026:CCP,GolWul:2026:SSM}.
It will be interesting to observe how the EMSL, or an improved variant thereof, interacts with the discontinuous nonlinearity introduced by elasto-plasticity as well.
The hyperelastic problem setting investigated here, which results in moderately nonlinear behaviour and relatively predictable qualitative behaviour is perhaps especially well-suited to the estimate-then-linearise approach taken by EMSL. Modifications might be necessary to account for the onset and evolution of plastic flow.

In particular, the estimation of cluster-wise expected average strains in EMSL could be improved. So far, we use a simple interpolation from parameter space to reduced space. In dramatically nonlinear problems, this interpolation might lose robustness and accuracy. Since the purpose of this work was the generalisation of strain-space methods to solid-mechanical problems beyond computational homogenisation rather than the further development of individual MOR techniques, no investigations to this end were performed here. 

While much work remains to be done, the dramatic runtime reductions strain-space MOR techniques seem to be able to facilitate (while retaining high levels of accuracy) might open up a wide range of possible engineering applications. These might include the control of flexible actuators, costly optimisation problems, and the fitting of material models, just to name a few.


    
\section*{Acknowledgements}

We would like to thank Ahmad Awad for suggesting the example problem investigated here, Dong Zhao for his contributions to the implementation of E3C, and Stephan Wulfinghoff and Hauke Goldbeck for their helpful suggestions regarding E3C and the further development of EMSL, as well as helpful suggestions regarding this manuscript.
We also gratefully acknowledge funding from DFG-Grant TRR 375 511263698. We used OpenAI's ChatGPT 5.3 to proof-read this manuscript and to shade Tables~\ref{tab:ECSW_error},~\ref{tab:ECM_error},~\ref{tab:E3C_error} and~\ref{tab:EMSL_error}.

\bibliography{sources}

\begin{appendices}

\section{Timing data}\label{a:timing}

\begin{table}[H]
    \centering
    \begin{tblr}{
    colspec={cc|ccccccccc},
    }   
        & & \SetCell[c=9]{c} $|E_\text{m}|$ & \\
        & & 1 & 2 & 5 & 10 & 20 & 30  & 50 & 75 & 100  \\
        \hline
        \SetCell[r=6]{l} $d$ 
 & 5 & 0.006084 & $\times$  & $\times$  & 0.05071 & 0.1004 & 0.1509 & 0.2502 & 0.3671 & 0.4881 \\
 & 7 & $\times$  & $\times$  & 0.02802 & 0.05159 & 0.1002 & 0.1511 & 0.2508 & 0.3752 & 0.4965 \\
 & 10 & $\times$  & $\times$  & 0.02735 & 0.05216 & 0.1021 & 0.1515 & 0.2504 & 0.3752 & 0.5421 \\
 & 15 & $\times$  & $\times$  & $\times$  & 0.05633 & 0.1060 & 0.1568 & 0.2543 & 0.3831 & 0.5122 \\
 & 20 & $\times$  & $\times$  & $\times$  & 0.06491 & 0.1063 & 0.1590 & 0.2584 & 0.3873 & 0.5260 \\
 & 30 & $\times$  & $\times$  & 0.1575 & $\times$  & 0.1080 & $\times$  & 0.2644 & 0.3874 & 0.5146 \\
    \end{tblr}
    \caption{Relative runtime ($\%$) obtained using displacement-space ECSW models with different $d$ and $|E_\text{m}|$, relative to original FE simulation.}
    \label{tab:ECSW_time}
\end{table}

\begin{table}[H]
\footnotesize
    \centering
    \begin{tblr}{
    colspec={cc|ccccccccc},
    }   
        & & \SetCell[c=9]{c} $|H|$ & \\
        & & 1 & 2 & 5 & 10 & 20 & 30  & 50 & 75 & 100  \\
        \hline
        \SetCell[r=6]{l} $d$ 
 & 5 & 0.000694 & $\times$ & 0.003012 & 0.004950 & 0.008985 & 0.01346 & 0.02185 & 0.03197 & 0.04228 \\
 & 7 & $\times$ & 0.001211 & $\times$ & 0.004895 & 0.008999 & 0.01352 & 0.02181 & 0.03204 & 0.04258 \\
 & 10 & $\times$ & 0.001244 & 0.003383 & 0.004915 & 0.009224 & 0.01357 & 0.02209 & 0.03248 & 0.04260 \\
 & 15 & $\times$ & $\times$ & $\times$ & 0.005954 & 0.009396 & 0.01395 & 0.02232 & 0.03291 & 0.04356 \\
 & 20 & $\times$ & $\times$ & $\times$ & $\times$ & 0.009609 & 0.01373 & 0.02212 & 0.03284 & 0.04366 \\
 & 30 & $\times$ & $\times$ & $\times$ & $\times$ & 0.009966 & 0.01433 & 0.02268 & 0.03301 & 0.04388 \\
    \end{tblr}
    \caption{Relative runtime ($\%$) obtained using strain-space ECM models with different $d$ and $|H|$, relative to original FE simulation.}
    \label{tab:ECM_time}
\end{table}

\begin{table}[H]
\footnotesize
    \centering
    \begin{tblr}{
    colspec={cc|ccccccccc},
    }   
        & & \SetCell[c=9]{c} $|H|$ & \\
        & & 1 & 2 & 5 & 10 & 20 & 30  & 50 & 75 & 100  \\
        \hline
        \SetCell[r=6]{l} $d$ 
 & 5 & 0.0007829 & 0.001984 & 0.002892 & 0.006186 & 0.008973 & 0.01326 & 0.02235 & 0.03172 & 0.04194 \\
 & 7 & $\times$ & $\times$ & 0.002505 & 0.004710 & 0.008910 & 0.01329 & 0.02162 & 0.03192 & 0.04223 \\
 & 10 & $\times$ & 0.001183 & 0.002307 & 0.004435 & 0.008516 & 0.01501 & 0.02175 & 0.03240 & 0.04243 \\
 & 15 & $\times$ & $\times$ & $\times$ & 0.004946 & 0.009453 & 0.01391 & 0.02224 & 0.03265 & 0.04289 \\
 & 20 & $\times$ & $\times$ & $\times$ & 0.004928 & 0.009358 & 0.01465 & 0.02220 & 0.03276 & 0.04312 \\
 & 30 & $\times$ & $\times$ & $\times$ & $\times$ & 0.009337 & 0.01396 & 0.02223 & 0.03301 & 0.04364 \\
    \end{tblr}
    \caption{Relative runtime ($\%$) obtained using strain-space E3C models with different $d$ and $|H|$, relative to original FE simulation.}
    \label{tab:E3C_time}
\end{table}

\begin{table}[H]
\footnotesize
    \centering
    \begin{tblr}{
    colspec={cc|ccccccccc},
    }   
        & & \SetCell[c=9]{c} $|H|$ & \\
        & & 1 & 2 & 5 & 10 & 20 & 30  & 50 & 75 & 100 \\
        \hline
        \SetCell[r=6]{l} $d$ 
 & 5 & 0.0002288 & 0.0003312 & 0.0007826 & 0.001449 & 0.002847 & 0.004178 & 0.006753 & 0.01001 & 0.01312 \\
 & 7 & 0.0002161 & 0.0003651 & 0.0008103 & 0.001507 & 0.002904 & 0.004376 & 0.006825 & 0.01019 & 0.01341 \\
 & 10 & 0.0002332 & 0.0004283 & 0.0008186 & 0.001557 & 0.003010 & 0.004496 & 0.007368 & 0.01086 & 0.01425 \\
 & 15 & 0.0002577 & 0.0004129 & 0.0009094 & 0.001645 & 0.003185 & 0.004843 & 0.008036 & 0.01159 & 0.01507 \\
 & 20 & 0.0002653 & 0.0004270 & 0.0009092 & 0.001714 & 0.003369 & 0.004858 & 0.008242 & 0.01291 & 0.01618 \\
 & 30 & 0.0002849 & 0.0005280 & 0.001112 & 0.001979 & 0.003845 & 0.005744 & 0.009691 & 0.01448 & 0.01810 \\
    \end{tblr}
    \caption{Relative runtime ($\%$) obtained using strain-space EMSL models with different $d$ and $|H|$, relative to original FE simulation.}
    \label{tab:EMSL_time}
\end{table}

\end{appendices}


\end{document}